\documentclass[journal]{IEEEtran}
\usepackage{amsmath,amsfonts}
\usepackage{algorithm}
\usepackage{array}
\usepackage[caption=false,font=normalsize,labelfont=sf,textfont=sf]{subfig}
\usepackage{textcomp}
\usepackage{stfloats}
\usepackage{url}
\usepackage{verbatim}
\usepackage{graphicx}
\usepackage{cite}
\def\hidephoto{true}

\usepackage{enumitem}

\usepackage{amssymb}

\usepackage{booktabs}

\usepackage{algpseudocode}

\usepackage{bm}

\PassOptionsToPackage{svgnames}{xcolor}
\usepackage{xcolor}

\usepackage{mdframed}

\newcommand{\category}[1]{\texttt{#1}}

\newcommand{\userStudyMethod}[1]{\textit{#1}}
\newcommand{\dataset}[1]{\textit{#1}}

\title{Viewpoint Recommendation for Point Cloud Labeling through Interaction Cost Modeling}

\author{
    Yu~Zhang,
    Xinyi~Zhao,
    Chongke~Bi,
    Siming~Chen%
    \thanks{
        Yu Zhang is with School of Data Science, Fudan University, Shanghai 200433, China, and also with the Department of Computer Science, University of Oxford, OX1 2JD Oxford, U.K. (e-mail: yuzhang94@outlook.com).
    }
    \thanks{
        Xinyi Zhao is with School of Data Science, Fudan University, Shanghai 200433, China (e-mail: zxy1337445805@gmail.com).
    }
    \thanks{
        Chongke Bi is with College of Intelligence and Computing, Tianjin University, Tianjin 300072, China (e-mail: bichongke@tju.edu.cn).
    }
    \thanks{
        Siming Chen is with School of Data Science, Fudan University, Shanghai 200433, China, and also with the Shanghai Key Laboratory of Data Science, Shanghai 200433, China (e-mail: simingchen@fudan.edu.cn).
    }
    \thanks{
        Yu Zhang and Xinyi Zhao contributed equally to this work.
        Siming Chen is the corresponding author.
    }
    \thanks{
        This article has supplementary downloadable material available at https://doi.org/10.1109/TVCG.2024.3376951, provided by the authors.
    }

    \thanks{
        © 2024 IEEE.
        This is the author's version of the article that has been accepted in IEEE Transactions on Visualization and Computer Graphics.
        Personal use of this material is permitted.
        Permission from IEEE must be obtained for all other uses, in any current or future media, including reprinting/republishing this material for advertising or promotional purposes, creating new collective works, for resale or redistribution to servers or lists, or reuse of any copyrighted component of this work in other works.
    }
}

\IEEEpubid{1077-2626~\copyright~2024 IEEE}

\begin{document}

\ifx\hidemain\undefined

  \maketitle

  \begin{abstract}
    Semantic segmentation of 3D point clouds is important for many applications, such as autonomous driving.
To train semantic segmentation models, labeled point cloud segmentation datasets are essential.
Meanwhile, point cloud labeling is time-consuming for annotators, which typically involves tuning the camera viewpoint and selecting points by lasso.
To reduce the time cost of point cloud labeling, we propose a viewpoint recommendation approach to reduce annotators' labeling time costs.
We adapt Fitts' law to model the time cost of lasso selection in point clouds.
Using the modeled time cost, the viewpoint that minimizes the lasso selection time cost is recommended to the annotator.
We build a data labeling system for semantic segmentation of 3D point clouds that integrates our viewpoint recommendation approach.
The system enables users to navigate to recommended viewpoints for efficient annotation.
Through an ablation study, we observed that our approach effectively reduced the data labeling time cost.
We also qualitatively compare our approach with previous viewpoint selection approaches on different datasets.

  \end{abstract}


  \section{Introduction}
\label{sec:introduction}

\IEEEPARstart{P}{oint} cloud data obtained from light detection and ranging (LiDAR) sensors have found wide-ranging applications, such as autonomous driving, medical imaging, and mixed reality.
Many of these applications rely on semantic segmentation of point cloud data.
Training deep learning models to perform semantic segmentation requires point cloud datasets with point-level labels.

Constructing 3D point cloud datasets with point-level labels, such as the SemanticKITTI dataset~\cite{Behley2019SemanticKITTI}, is labor-intensive.
The annotator needs to iteratively identify objects to label, adjust the camera viewpoint, select points, and assign labels.
Techniques have thus been introduced to improve the efficiency of 3D point cloud labeling, such as active learning~\cite{Yi2016Scalable} and intelligent selection~\cite{Chen2020LassoNet}.

Recommending suitable viewpoints may reduce the time cost for annotators to adjust the viewpoint and improve the efficiency of data labeling.
For example, when annotating a vehicle in a point cloud, it may reduce the effort of selecting the points belonging to the vehicle by recommending a viewpoint where the vehicle is easily separable from the remaining points.
Viewpoint recommendation or selection has long been studied for 3D visualization.
Meanwhile, existing viewpoint selection approaches, such as viewpoint entropy~\cite{Vazquez2001Viewpoint}, are not designed to optimize labeling time cost.

This work proposes a viewpoint recommendation approach to reduce the time cost of point cloud labeling.
We aim to reduce the time the annotator spends on tuning the camera viewpoint and selecting points with lasso selection.
We adapt Fitts' law~\cite{Fitts1992Information} to model and optimize the time cost of lasso selection in point cloud labeling (Sec.~\ref{sec:approach}).
We develop a point cloud labeling system that integrates our viewpoint recommendation approach (Sec.~\ref{sec:data-labeling-system}).
An ablation study suggests that our approach reduces labeling time cost (Sec.~\ref{sec:ablation-study}).
We also qualitatively compare our approach with other viewpoint recommendation approaches (Sec.~\ref{sec:comparative-analysis}).
We discuss future work directions, particularly visualization for routine tasks (Sec.~\ref{sec:discussion}).

In summary, this paper has the following contributions:

\begin{itemize}[leftmargin=*]
    \item We propose a time cost model for lasso selection in 2D scatter plots based on Fitts' law.

    \item We propose a viewpoint recommendation approach for point cloud labeling based on the time cost model.

    \item We have developed a point cloud labeling system that integrates the viewpoint recommendation approach.
          We have examined its effectiveness through an ablation study and a qualitative comparison.
\end{itemize}

  \IEEEpubidadjcol
  \section{Related Work}

This section introduces point cloud labeling techniques and viewpoint selection methods.

\subsection{Point Cloud Labeling}

Techniques to improve the efficiency of point cloud labeling include assigning default labels, active learning, and intelligent selection.

\textbf{Default labels} can be assigned to unlabeled points to reduce the efforts of annotators.
The default labels may utilize the labels of other points in the neighborhood~\cite{Monica2017Multi} or labels assigned to other similar point clouds~\cite{Yi2016Scalable}.
For time-varying point clouds, the default labels may be based on labels for previous frames~\cite{Wang2019LATTE,Zimmer20193D,Li2020SUSTech}.

\textbf{Active learning} is a family of techniques that samples unlabeled data points for the annotator to label.
It has been used to reduce the labor of semantic segmentation for point clouds.
Yi et al. propose an active learning framework that models and optimizes annotation time cost~\cite{Yi2016Scalable}.
Luo et al. use margin sampling to sample supervoxels in point clouds for labeling~\cite{Luo2018Semantic}.
Shao et al. use graph reasoning to sample supervoxels with spatial and structural diversity~\cite{Shao2022Active}.

\textbf{Point selection} is a fundamental interaction in point cloud labeling.
Conventionally, lasso selection for 3D point clouds is implemented as a cylinder selection in 3D space.
Techniques have been developed to improve the efficiency of point selection.
Yu et al. propose two lasso selection techniques that utilize the spatial structure of point clouds~\cite{Yu2012Efficient}.
Yu et al. develop a family of context-aware selection techniques that infer the user's intention from the user's actual selection and utilize point density information to refine the selection~\cite{Yu2016CAST}.
Fan et al. train a convolutional neural network with user study data to estimate the intended selection in 2D scatter plots given the user's bounding box selection~\cite{Fan2018Fast}.
Chen et al. trained a hierarchical neural network with user selection data to refine the user's lasso selection regions in point cloud~\cite{Chen2020LassoNet}.

Similar to this thread of work, we aim to improve the efficiency of point cloud interaction.
We focus on viewpoint selection, which is orthogonal to the prior work.
Thus, our method may be used together with the prior work.

\subsection{Viewpoint Selection}
\label{sec:viewpoint-selection}

The appearance of a 3D model heavily depends on the viewpoint.
Viewpoint selection is thus critical in rendering 3D models.
Generally, a viewpoint can be parameterized as a matrix that projects a point in the 3D space to a point in the 2D screen.
Techniques have been proposed in the literature for viewpoint selection.

\textbf{Principal component analysis} (PCA) is commonly used for normalizing the pose of 3D data~\cite{Li2021Closer}.
For 3D points, using the third principal component as the viewing direction maximizes the spatial variance of the points on a 2D screen.
Thus, using the third principal component as the viewpoint may minimize the occlusion of interesting features~\cite{Kim2013new}.

\textbf{Entropy-based} viewpoint selection aims to maximize the amount of information that the user can obtain within the viewpoint.
Vázquez et al. propose a viewpoint entropy metric for mesh data that favors viewpoints where surfaces are equally visible~\cite{Vazquez2001Viewpoint}.
Takahashi et al. extend the viewpoint entropy metric to volumetric data by decomposing volume into feature subvolumes, computing and weighing the viewpoint entropy of feature subvolumes~\cite{Takahashi2005Feature}.
Bordoloi and Shen propose an entropy-based metric that assigns a noteworthiness score to each voxel based on the transfer function and data distribution, and selects viewpoints where the noteworthiness of voxels is proportional to their visibility~\cite{Bordoloi2005View}.
Viola et al. propose a viewpoint mutual information metric to measure the dependence between the viewpoint and pre-segmented objects in volumetric data, and recommend viewpoints given the user's object of interest~\cite{Viola2006Importance}.
The viewpoint mutual information is also extended to mesh data~\cite{Feixas2009Unified}.
Ji and Shen use the opacity entropy, color entropy, and curvature information in the rendered image of volumetric data for viewpoint selection~\cite{Ji2006Dynamic}.
Bramon et al. suggest selecting the viewpoint that maximizes the mutual information between the intensity of volumetric data and the rendered image colors~\cite{Bramon2013Information}.
Marsaglia et al. propose viewpoint quality metrics based on data entropy, shading entropy, and depth entropy and investigate mechanisms for combining metrics~\cite{Marsaglia2021Entropy}.
Based on these metrics, Marsaglia et al. propose an efficient in situ camera placement approach for isosurfaces to optimize the viewpoint quality while minimizing execution time~\cite{Marsaglia2022Automatic}.

\textbf{Other features} in addition to entropy have also been used for viewpoint selection.
Kamada and Kawai suggest selecting viewpoints that minimize the number of degenerated surfaces~\cite{Kamada1988Simple}.
Weinshall and Weman propose the metrics of view likelihood and stability~\cite{Weinshall1997View}.
Stoev and Straßer propose selecting viewpoints that maximize the projected area and depth of the displayed scene~\cite{Stoev2002Case}.
Podolak suggests selecting viewpoints that minimize the symmetry of the displayed model~\cite{Podolak2006Planar}.

\textbf{Learning-based} methods learn from user preferences on viewpoints.
Vieira et al. extracts object-space and image-space features for viewpoint, use support vector machines to learn from user preference, and use the learned model to classify viewpoint as good or bad~\cite{Vieira2009Learning}.
Secord et al. extend this learning-based approach to more viewpoint features and fit regression models of viewpoint goodness with a large user study data~\cite{Secord2011Perceptual}.
While the viewpoint preference model may learn from the user, other information sources, such as web images~\cite{Liu2011Web} and publication images~\cite{Tao2016Similarity}, may also be used for preference learning.

Each viewpoint selection method reflects a measure of viewpoint goodness.
Previous work on viewpoint selection focuses on data analysis scenarios, where the goal is to help users understand the 3D data efficiently.

Unlike the previous work, our usage scenario is data labeling.
The goal is thus to help users label the 3D data efficiently.
The difference in the underlying objective leads to different measures of viewpoint goodness.
Our metric of a good viewpoint is to minimize the time cost for users to select points, which differs from the optimization objectives in the literature.
A viewpoint where a 3D model's features overlap may be bad for data analysis but may be suitable for data labeling, as the model may occupy a smaller area in the 2D screen and is easier to select with lasso selection.
Section~\ref{sec:comparative-analysis} qualitatively compares ours and prior viewpoint selection approaches.

  \section{Background on Point Cloud Labeling}

For point cloud labeling, we introduce common types of interactions and the labeling process with lasso selection to motivate our viewpoint recommendation approach.

\subsection{Labeling Interactions}

We examine the interaction design in 13 point cloud data labeling systems in Table~\ref{table:labeling-interaction-survey}.
We observe four types of labeling interactions in these systems:

\begin{itemize}[leftmargin=*]
    \item \textbf{Cuboid:}
          The annotator creates a bounding cuboid, typically axes-aligned, by specifying three control points.

    \item \textbf{Lasso:}
          The annotator creates a closed polygon by freeform drawing.

    \item \textbf{Landmark:}
          The annotator creates a landmark point by clicking.

    \item \textbf{Polyline:}
          The annotator creates a polyline by freeform drawing or clicking to specify control points.
\end{itemize}

We focus on lasso selection as it is common in point cloud labeling for semantic segmentation.
While it is less frequently implemented than bounding cuboid, lasso selection allows precise selection of points and thus may suit more usage scenarios.

\begin{table}[ht]
    \centering
    \caption{
        Labeling interactions in point cloud labeling systems.
    }
    \label{table:labeling-interaction-survey}
    \begin{tabular}{llr}
        \toprule
        \textbf{Interaction} & \textbf{System}                                                                                                                                                                         & \textbf{Count}                            \\
        \midrule
        Cuboid               & \cite{GirardeauMontaut2003CloudCompare,SageMaker2017Use,Zimmer20193D,Wang2019LATTE,Sekachev2020CVAT,Li2020SUSTech,Sager2022labelCloud,Supervisely2023Supervisely,Foundation2023Xtreme1} & 9/13 \rule[-4pt]{0pt}{0pt}                \\
        \hline
        Lasso                & \cite{GirardeauMontaut2003CloudCompare,SageMaker2017Use,Behley2019SemanticKITTI,Zhang2022OneLabeler,Supervisely2023Supervisely,Foundation2023Xtreme1}                                   & 6/13 \rule{0pt}{8pt}\rule[-4pt]{0pt}{0pt} \\
        \hline
        Landmark             & \cite{Monica2017Multi,Kontogianni2023Interactive,Supervisely2023Supervisely}                                                                                                            & 3/13 \rule{0pt}{8pt}\rule[-4pt]{0pt}{0pt} \\
        \hline
        Polyline             & \cite{Foundation2023Xtreme1}                                                                                                                                                            & 1/13 \rule{0pt}{8pt}                      \\
        \bottomrule
    \end{tabular}
\end{table}

\subsection{Labeling Process with Lasso Selection}

Consider the scenario where the annotator aims to label a vehicle in a point cloud obtained with LiDAR sensors.
In this scenario, the annotator needs first to identify the vehicle.
Then, the annotator needs to move the camera so that other points do not occlude the points corresponding to the vehicle.
Finally, the annotator needs to draw a lasso polygon to assign labels to the points corresponding to the vehicle.
The annotator may perform this process iteratively to label multiple objects.

\begin{figure}[!htbp]
    \centering
    \includegraphics[width=0.8\linewidth]{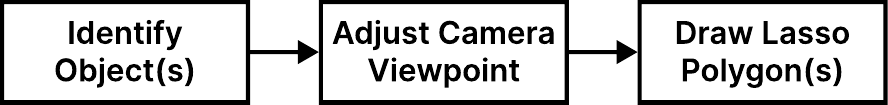}
    \caption{
        \textbf{Steps of using lasso selection for semantic segmentation:}
        (1) Identify object(s) to label in the 3D scene.
        (2) Adjust the camera viewpoint to focus on the objects to be labeled and avoid overlapping the objects with points of other categories.
        (3) Draw lasso polygon(s) to select points and assign labels.
    }
    \label{fig:lasso-labeling-steps}
\end{figure}

As summarized in Fig.~\ref{fig:lasso-labeling-steps}, labeling a point cloud with lasso selection typically involves identifying objects to label, adjusting the viewpoint, and drawing a lasso polygon.

We aim to reduce the time cost of labeling with lasso selection.
We focus on reducing the time cost of the latter two steps in Fig.~\ref{fig:lasso-labeling-steps}, i.e., adjusting viewpoint and drawing lasso polygons.
Our approach is to recommend viewpoints to annotators to reduce the time required to adjust the camera viewpoint.
The recommended viewpoints are optimized to minimize the time cost of drawing lasso polygons.

  \section{Viewpoint Recommendation Approach}
\label{sec:approach}

Our goal is to reduce the time cost of semantic segmentation.
Thus, we recommend viewpoints that reduce the time cost to select points belonging to the same object.

This section first introduces background on Fitts' law for modeling mouse movement time costs (Sec.~\ref{sec:fitts-law}).
Then, we describe adapting Fitts' law to model the time cost of lasso selection in point cloud labeling (Sec.~\ref{sec:lasso-time-cost}).
For a given viewpoint, a point cloud is displayed as a 2D scatter plot.
We use Fitts' law to model the time cost of lasso selection in the 2D scatter plot.
With the time cost model, we use grid search to optimize viewpoint (Sec.~\ref{sec:viewpoint-optimization}).
Finally, we reflect on the assumptions and considerations (Sec.~\ref{sec:assumptions-considerations}).

\begin{figure}[!htbp]
    \centering
    \includegraphics[width=\linewidth]{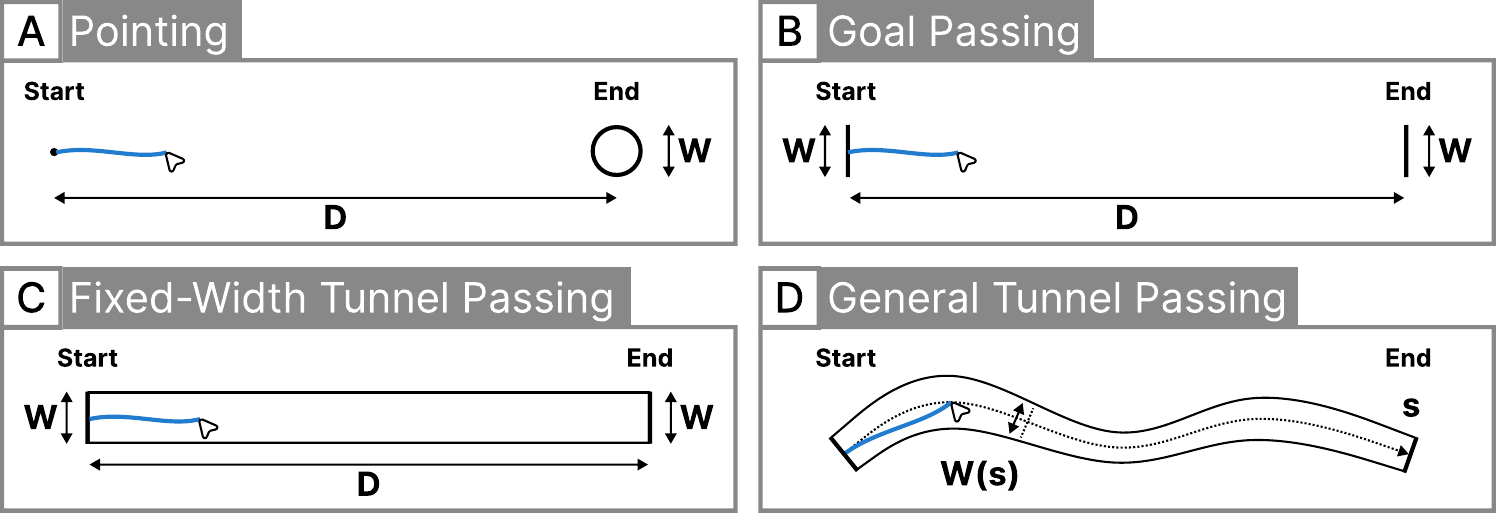}
    \caption{
        \textbf{Fitts' law for mouse movement tasks:}
        (A) For pointing, $T = a_p + b_p log_2 (\frac{D}{W} + 1)$.
        (B) For goal passing, $T = a_g + b_g log_2 (\frac{D}{W} + 1)$.
        (C) For fixed-width tunnel passing, $T = a_s + b_s \frac{D}{W}$.
        (D) For general tunnel passing, $T = a_s + b_s \int_c \frac{ds}{W(s)}$.
    }
    \label{fig:fitts-law}
\end{figure}

\subsection{Background of Fitts' Law}
\label{sec:fitts-law}

Fitts' law~\cite{Fitts1992Information} is widely used to model the time cost of mouse pointing tasks.
Accot and Zhai~\cite{Accot1997Fitts} extended Fitts' law to trajectory-based interaction tasks.
The following introduces the time cost models of four mouse movement tasks derived from Fitts' law: pointing, goal passing, fixed-width tunnel passing, and general tunnel passing.

\textbf{Pointing task:}
Conventionally, Fitts' law focuses on pointing tasks where the user needs to move the mouse to a target, as shown in Fig.~\ref{fig:fitts-law}(A).
According to the law, the mouse movement time cost $T$ can be modeled as:

\begin{equation}\label{eq:object-pointing}
    T = a_p + b_p log_2 (\frac{D}{W} + 1)
\end{equation}

\begin{itemize}[leftmargin=*]
    \item $D$ is the distance between the start and target points.

    \item $W$ is the target size.

    \item $a_p$ and $b_p$ are constants that can be estimated with user experiments.
          Normally, $b_p > 0$, as the time cost should be higher for longer distances and smaller targets.

    \item $ID_p = log_2 (\frac{D}{W} + 1)$ is commonly referred to as the index of difficulty of the task.
\end{itemize}

\textbf{Goal passing task:}
In this task, the user needs to move the mouse to pass the start and end points with width $W$, as shown in Fig.~\ref{fig:fitts-law}(B).
$D$ is the distance from start to end.
Accot and Zhai's experiment~\cite{Accot1997Fitts} shows that the movement time between start and end can be modeled with the same function in Equation~\ref{eq:object-pointing}:

\begin{equation}\label{eq:t-goal-passing}
    T = a_g + b_g log_2 (\frac{D}{W} + 1)
\end{equation}

\begin{itemize}[leftmargin=*]
    \item $ID_g = log_2 (\frac{D}{W} + 1)$ is the index of difficulty of the task.
\end{itemize}

\textbf{Fixed-width tunnel passing task:}
In this task, the user needs to move the mouse to pass the start and end points with width $W$.
Compared with the goal passing task, the tunnel passing task requires that the mouse movement be constrained within the tunnel of width $W$, as shown in Fig.~\ref{fig:fitts-law}(C).
Accot and Zhai~\cite{Accot1997Fitts} suggest that the tunnel passing task can be seen as sequentially carrying out $N$ goal passing tasks with $N \to \infty$.
The distance from start to end in each goal-passing task is $\frac{D}{N}$.
Under this perspective, the time cost of the tunnel passing task can be seen as an integral of the time costs of goal passing tasks.
Through experiments, Accot and Zhai~\cite{Accot1997Fitts} verified that the time cost of the tunnel passing task can be modeled as:

\begin{equation}\label{eq:t-fixed-width-tunnel}
    T = a_s + b_s \frac{D}{W}
\end{equation}

\textbf{General tunnel passing task:}
More generally, Accot and Zhai~\cite{Accot1997Fitts} suggest that the time cost for a complex tunnel passing task can be modeled as:

\begin{equation}\label{eq:t-general-tunnel}
    T = a_s + b_s \int_c \frac{ds}{W(s)}
\end{equation}

\begin{itemize}[leftmargin=*]
    \item $c$ is the medial axis of the tunnel, which is a curved path.

    \item $W(s)$ is the width of the tunnel at position $s$ along the curvilinear axis formed by $c$.

    \item $a_s$ and $b_s$ are constants that can be estimated with user experiments.
          Normally, $b_s > 0$, as the time cost should be higher for longer distances and narrower tunnels.

    \item $ID_s = \int_c \frac{ds}{W(s)}$ is the index of difficulty of the task.
\end{itemize}

Note that the time cost of various mouse movement tasks can be normalized to a linear function $T = a + b ID$, where $ID$ is referred to as the task's index of difficulty.
For the different tasks, $ID$ usually takes different forms, and the constants $a$ and $b$ in the equations may take different values.

In the following, we demonstrate adapting the equations above to model the lasso selection time cost.

\subsection{Modeling Lasso Selection Time Cost}
\label{sec:lasso-time-cost}

In point cloud labeling, lasso selection is a constrained mouse movement task.
To avoid misclassifying points, the annotator needs to ensure the lasso polygon encloses and only encloses the points belonging to the target object.

Given a viewpoint, a point cloud is projected onto a 2D plane and displayed as a 2D scatter plot.
In the following, we model the time cost of lasso selection in 2D scatter plots.

\begin{mdframed}
    Our modeling is based on the geometric intuition that \textbf{tunnel passing is equivalent to lasso selection in infinitely dense point clouds}.
\end{mdframed}

\begin{figure}[!htbp]
    \centering
    \includegraphics[width=\linewidth]{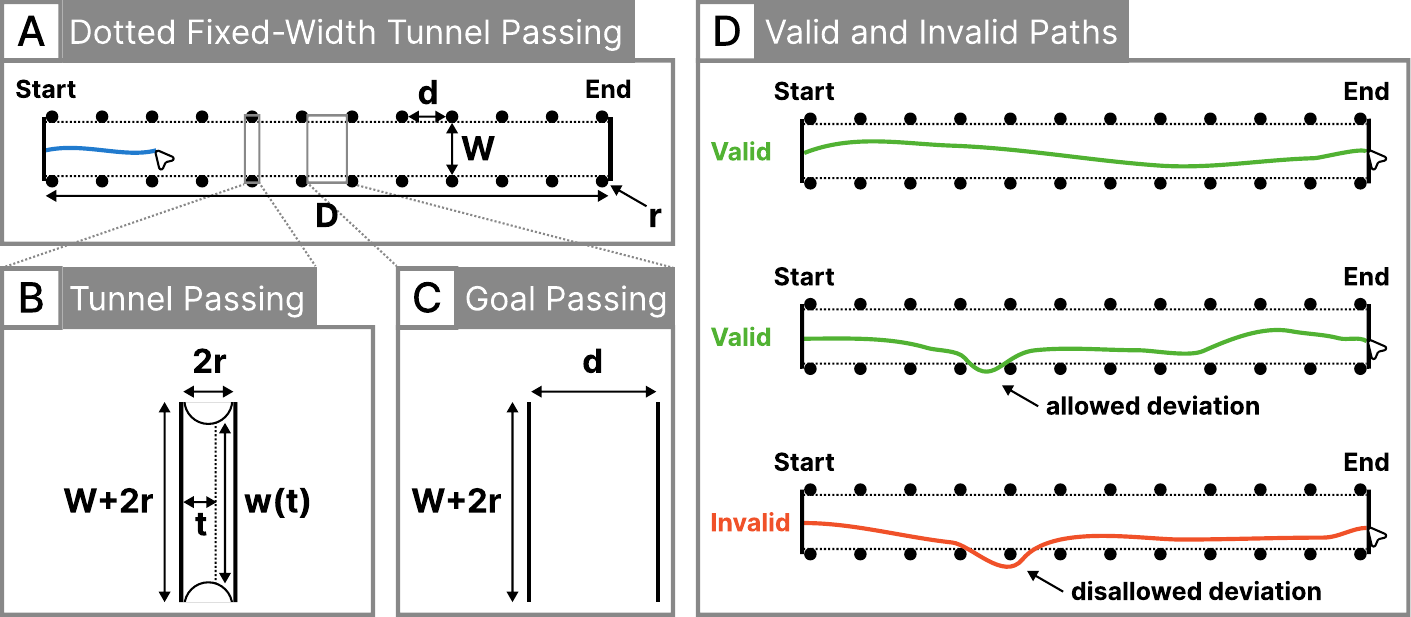}
    \caption{
        \textbf{Dotted fixed-width tunnel passing:}
        (A) A dotted fixed-width tunnel passing task has parameters $D$, $W$, $r$, and $d$.
        We regard that a dotted fixed-width tunnel passing task can be decomposed into two types of subtasks: curved tunnel passing subtasks and goal passing subtasks.
        (B) A curved tunnel passing subtask has parameters $W$ and $r$.
        The tunnel width at $t \in [0, 2r]$ is $w(t) = W + 2r - 2\sqrt{r^2 - (t-r)^2}$.
        (C) A goal passing subtask has parameters $W$, $r$, and $d$.
        (D) Examples of valid and invalid mouse movement paths for dotted fixed-width tunnel passing.
        The dotted fixed-width tunnel passing task poses less restrictions than the fixed-width tunnel passing task in Fig.~\ref{fig:fitts-law}(C).
        The mouse movement path is allowed to exceed the tunnel but not allowed to enclose any point.
    }
    \label{fig:dotted-tunnel}
\end{figure}

\begin{figure*}[!htbp]
    \centering
    \includegraphics[width=\linewidth]{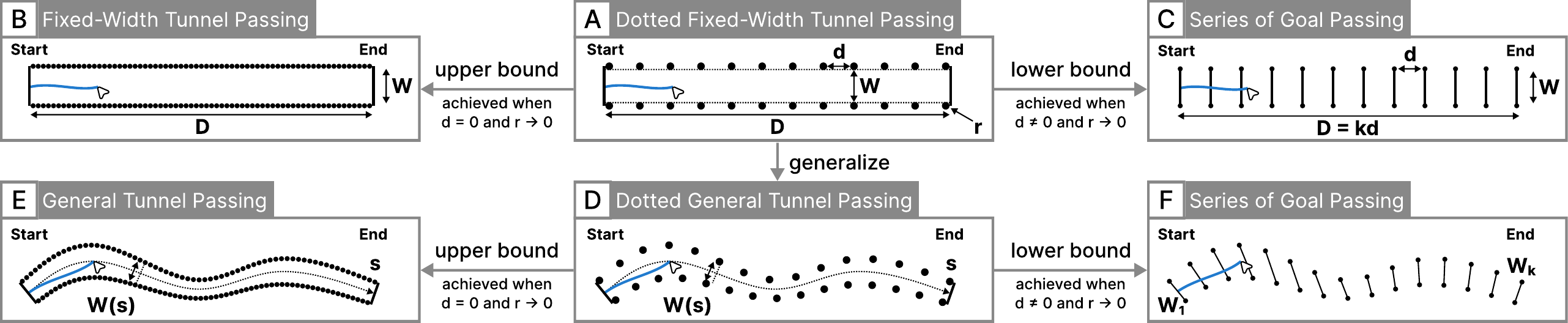}
    \caption{
        \textbf{Upper and lower bounds of the dotted fixed-width tunnel passing:}
        (A) The original dotted fixed-width tunnel passing task has parameters $D \in \mathbb{R}^+$, $W \in \mathbb{R}^+$, $r \in \mathbb{R}^+$, and $d \in \mathbb{R}_{\geq 0}$.
        Let $ID(D, W, r, d)$ denote the index of difficulty of this task, which is a function with arguments $D$, $W$, $r$, and $d$.
        (B) The upper bound of $ID(D, W, r, d)$ is $\frac{D}{W}$.
        The upper bound is achieved when $d = 0$ and $r \rightarrow 0^+$.
        In this case, geometrically, the dotted fixed-width tunnel passing task is equivalent to a fixed-width tunnel passing task (see Fig.~\ref{fig:fitts-law}(C)).
        (C) The lower bound of $ID(D, W, r, d)$ is $mk log_2(\frac{D}{kW} + 1)$.
        The lower bound is achieved when $d \neq 0$ and $r \rightarrow 0^+$.
        In this case, geometrically, the dotted fixed-width tunnel passing task is equivalent to a series of $k$ goal passing tasks (see Fig.~\ref{fig:fitts-law}(B)).
        (D) Compared to the dotted fixed-width tunnel passing task, the dotted general tunnel passing task no longer requires the points to be evenly distributed on two parallel line segments.
        (E) The dotted general tunnel passing task is geometrically equivalent to a general tunnel passing task (see Fig.~\ref{fig:fitts-law}(D)) when $d = 0$ and $r \rightarrow 0^+$.
        (F) The dotted general tunnel passing task is geometrically equivalent to a series of goal passing tasks task (see Fig.~\ref{fig:fitts-law}(B)) when $d \neq 0$ and $r \rightarrow 0^+$.
    }
    \label{fig:bounds}
\end{figure*}

\subsubsection{Dotted Fixed-Width Tunnel Passing Task}
\label{sec:dotted-tunnel-time-cost}

Before modeling the time cost of lasso selection, we start with a simpler case where we model the time cost of the dotted fixed-width tunnel passing task.
In this task, the user needs to draw a line that passes through a tunnel formed by two parallel dotted lines, as shown in Fig.~\ref{fig:dotted-tunnel}(A).
The mouse movement path has to suffice that the contour generated by closing the path does not enclose any point in the dotted lines, as shown in Fig.~\ref{fig:dotted-tunnel}(D).
We introduce the following notation for this task:

\begin{itemize}[leftmargin=*]
    \item $D$ is the distance between the start and the target.

    \item $W$ is the width of the tunnel.

    \item $r$ is the radius of the points on the dotted lines.

    \item $d$ is the gap between adjacent points in the dotted lines.
\end{itemize}

Without loss of generality, we assume that the start point and the target are aligned with the dotted lines.
In other words, there exists $k \in Z^+$ such that $D = 2r(k + 1) + dk$.
Equivalently, there exists $k \in Z^+$ such that:

\begin{equation}
    k = \frac{D - 2r}{d + 2r}
\end{equation}

In this case, dotted fixed-width tunnel passing can be decomposed into $2k + 1$ subtasks:

\begin{itemize}[leftmargin=*]
    \item $k + 1$ subtasks are curved tunnel passing tasks, as shown in Fig.~\ref{fig:dotted-tunnel}(B).
          Let $w(t)$ denote the width of the tunnel at the point along the medial axis that is $t$ distance away from the origin.
          We have $w(t) = W + 2r - 2\sqrt{r^2 - (r - t)^2}$.
          According to Equation~\ref{eq:t-general-tunnel}, the time cost of each subtask is $T_1 = a_s + b_s ID_1$.
          The index of difficulty of each subtask is $ID_1 = \int_c \frac{ds}{W(s)} = \int_0^{2r} \frac{dt}{W + 2r - 2\sqrt{r^2 - (r - t)^2}}$.
          As proved in
          \ifx\hideappendix\undefined
              Appendix~\ref{sec:proof-curved-tunnel-id},
          \else
              Appendix A.1 in the supplementary materials,
          \fi
          one can deduce that:

          \begin{equation}
              ID_1 = \frac{\frac{\pi}{2} + arcsin\frac{2r}{W + 2r}}{\sqrt{1 - \frac{4r^2}{(W + 2r)^2}}} - \frac{\pi}{2}
          \end{equation}

    \item The other $k$ of the subtasks are goal passing tasks with distance $d$ and goal width $W$, as shown in Fig.~\ref{fig:dotted-tunnel}(C).
          According to Equation~\ref{eq:t-goal-passing}, the time cost of each subtask is $T_2 = a_g + b_g ID_2$.
          The index of the difficulty of each subtask is:

          \begin{equation}
              ID_2 = log_2 (\frac{d}{W + 2r} + 1)
          \end{equation}
\end{itemize}

Adding up the time cost of all the subtasks, we derive the time cost of the dotted fixed-width tunnel passing task:

\begin{equation}\label{eq:t-dotted-fixed-width-tunnel-1}
    T = (k + 1)(a_s + b_s ID_1) + k(a_g + b_g ID_2)
\end{equation}

Let
$a_{c} = (k + 1)a_s + ka_g$,
$b_{c} = b_s$, and
$m = \frac{b_g}{b_s}$.
We rewrite Equation~\ref{eq:t-dotted-fixed-width-tunnel-1} to the following form:

\begin{equation}\label{eq:t-dotted-fixed-width-tunnel-2}
    T = a_c + b_c((k + 1)ID_1 + mkID_2)
\end{equation}

Thus, the index of difficulty of this dotted fixed-width tunnel passing task is:

\begin{equation}\label{eq:id-dotted-fixed-width-tunnel}
    ID = (k + 1)ID_1 + mkID_2
\end{equation}

In the following, we discuss the upper and lower bounds of $ID$ in Equation~\ref{eq:id-dotted-fixed-width-tunnel}.

\begin{itemize}[leftmargin=*]
    \item \textbf{Upper bound and corresponding limit:}
          \begin{itemize}
              \item \textbf{Bound:} $ID \leq \frac{D}{W}$
              \item \textbf{Limit:} $\lim_{r \rightarrow 0^+} (ID \rvert_{d=0}) = \frac{D}{W}$
          \end{itemize}
          \ifx\hideappendix\undefined
              Appendix~\ref{sec:upper-bound}
          \else
              Appendix A.2.2 in the supplementary materials
          \fi
          proves the bound and the limit.
          The index of difficulty of the fixed-width tunnel passing in Equation~\ref{eq:t-fixed-width-tunnel} is also $\frac{D}{W}$, the same as the bound and the limit.
          We interpret the correspondence as follows:

          \begin{itemize}
              \item Dotted fixed-width tunnel passing is \textit{no harder than} fixed-width tunnel passing.

              \item Dotted fixed-width tunnel passing \textit{equals} fixed-width tunnel passing \textit{when the points are small and dense}.
          \end{itemize}

          The correspondence is also intuitive from the geometry, as shown in Fig.~\ref{fig:bounds}(B).

    \item \textbf{Lower bound and corresponding limit:}
          \begin{itemize}
              \item \textbf{Bound:} $ID \geq mk log_2(\frac{D - 2r (k + 1)}{k (W + 2r)} + 1)$
              \item \textbf{Limit:} $\lim_{r \rightarrow 0^+} ID = mk log_2(\frac{D}{kW} + 1)$
          \end{itemize}
          \ifx\hideappendix\undefined
              Appendix~\ref{sec:lower-bound}
          \else
              Appendix A.2.3 in the supplementary materials
          \fi
          proves the bound and the limit.
          The index of difficulty of goal passing in Equation~\ref{eq:t-goal-passing} is $log_2(\frac{D}{W} + 1)$.
          When passing a series of $k$ goals evenly distributed along distance $D$, the total index of difficulty is $k log_2(\frac{D}{kW} + 1)$, the same as the bound (when $r$ is small) and the limit.
          We interpret the correspondence as follows:

          \begin{itemize}
              \item Dotted fixed-width tunnel passing is \textit{(approximately) no easier than} a series of goal passing.

              \item Dotted fixed-width tunnel passing \textit{equals} a series of goal passing \textit{when the points are small}.
          \end{itemize}

          The correspondence is also intuitive from the geometry, as shown in Fig.~\ref{fig:bounds}(C).
\end{itemize}

\subsubsection{Lasso Selection Task}

In the following, we generalize the upper bound of the time cost of the dotted fixed-width tunnel passing to lasso selection.
We first formally define the lasso selection task.

\begin{itemize}[leftmargin=*]
    \item $X$ is a list of points on the 2D plane:
          \begin{center}
              $X = (x_i)_{i=1}^n$ where $x_i \in \mathbb{R}^2$
          \end{center}

    \item $Y$ is a list of labels of the points in $X$:
          \begin{center}
              $Y = (y_i)_{i=1}^n$ where $y_i \in \{1, -1\}$
          \end{center}

    \item $X^+$ is the subset of $X$ with \category{positive} labels:
          \begin{center}
              $X^+ = \{x_i | x_i \in X \wedge y_i = 1\}$
          \end{center}

    \item $X^-$ is the subset of $X$ with \category{negative} labels:
          \begin{center}
              $X^- = \{x_i | x_i \in X \wedge y_i = -1\}$
          \end{center}
\end{itemize}

\textbf{Definition of a lasso selection task:}
Given points $X$ and labels $Y$, the user needs to draw a closed curve $c$ that encloses all points in $X^+$ and no point in $X^-$.
The lasso selection path $c$ has to suffice the following conditions:

\begin{itemize}[leftmargin=*]
    \item $c$ is a simple closed curve (i.e., a Jordan curve): $c$ is the image of a continuous map $\varphi: [0, 1] \rightarrow \mathbb{R}^2$ such that $\varphi(0) = \varphi(1)$ and $\varphi$ is injective on $[0, 1)$.

    \item $c$ encloses all points in $X^+$: For any $x^+ \in X^+$, the winding number of $\varphi$ around $x^+$ is not zero.

    \item $c$ encloses no points in $X^-$: For any $x^- \in X^-$, the winding number of $\varphi$ around $x^-$ is zero.
\end{itemize}

We refer to a lasso selection path $c$ that suffices the conditions above as \textit{valid}.
The points and labels $(X, Y)$ uniquely define a lasso selection task.

\textbf{Relation with the dotted fixed-width tunnel passing:}
Lasso selection can be seen as dotted general tunnel passing (Fig.~\ref{fig:bounds}(E)) that generalizes dotted fixed-width tunnel passing with the following two changes:

\begin{itemize}[leftmargin=*]
    \item \textbf{Point distribution:}
          For dotted fixed-width tunnel passing, the points are evenly distributed on the two parallel dotted lines along the tunnel.
          For lasso selection, the points can be arbitrarily distributed in the 2D plane.

    \item \textbf{Tunnel shape:}
          For dotted fixed-width tunnel passing, the tunnel is straight and fixed-width.
          For lasso selection, the point distribution is arbitrary, and the tunnel shape depends on the point distribution.
\end{itemize}

\textbf{Upper bound of the time cost:}
Similar to dotted fixed-width tunnel passing (Fig.~\ref{fig:bounds}(A)), for dotted general tunnel passing (Fig.~\ref{fig:bounds}(D)), it is geometrically intuitive that its upper bound is general tunnel passing (Fig.~\ref{fig:bounds}(E)) and its lower bound is a series of goal passing tasks (Fig.~\ref{fig:bounds}(F)).
Meanwhile, when estimating the lasso selection time cost, the tunnel parameterized as $(c, W(s))$ is not known in advance.

We refer to a tunnel $(c, W(s))$ as \textit{valid} when:

\begin{itemize}[leftmargin=*]
    \item A valid lasso selection path $c'$ exists inside the tunnel.
    \item No point in $X$ is inside the tunnel.
\end{itemize}

\begin{figure}[!htbp]
    \centering
    \includegraphics[width=\linewidth]{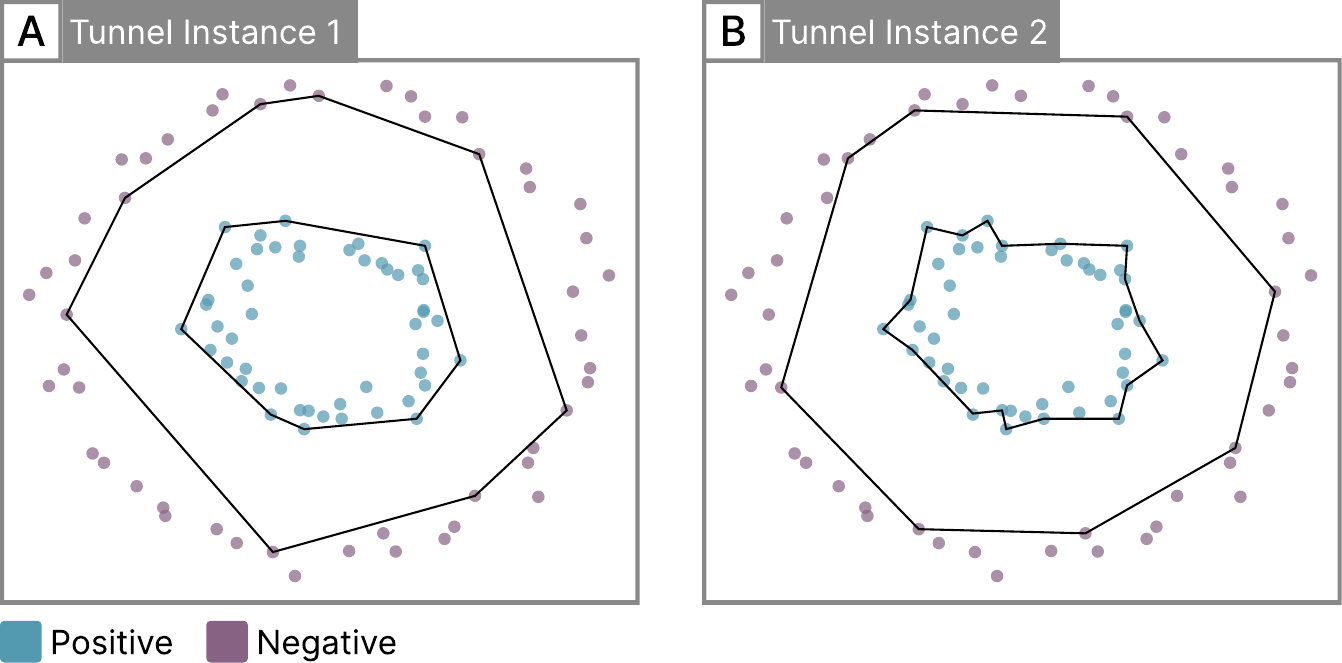}
    \caption{
        There can be multiple tunnels to estimate the lasso selection time cost for a given point cloud.
    }
    \label{fig:lasso-tunnels}
\end{figure}

Figure~\ref{fig:lasso-tunnels} shows that, given $(X, Y)$, the valid tunnel is not unique.
Let $T(c, W(s))$ denote the lasso selection time cost in a valid tunnel $\left(c, W(s)\right)$.

As shown in Fig.~\ref{fig:bounds}(E), an upper bound of lasso selection time cost in $\left(c, W(s)\right)$ is the time cost of the general tunnel passing in $\left(c, W(s)\right)$.
We minimize the upper bound to reduce the lasso selection time cost for the worst case.
Let $ID$ denote the index of difficulty of the lasso selection task defined by tunnel $\left(c, W(s)\right)$.
The following inequality holds for any valid tunnel $\left(c, W(s)\right)$.

\begin{equation}\label{eq:lasso-upper-bound}
    ID \leq \int_c \frac{ds}{W(s)}
\end{equation}

For any valid tunnel $\left(c, W(s)\right)$, the upper bound of lasso selection in $\left(c, W(s)\right)$ is also an upper bound of lasso selection given the task $(X, Y)$.
Furthermore, we can derive a tight upper bound of lasso selection time cost given $(X, Y)$ by finding the valid tunnel $\left(c, W(s)\right)$ that minimizes the right-hand side of Inequality~\ref{eq:lasso-upper-bound}.
Below, we introduce an algorithm that aims to minimize the right-hand side.

\subsubsection{Lasso Selection Time Cost Estimation Algorithm}
\label{sec:estimation-algorithm}

As discussed above, while it is hard to directly model the time cost of lasso selection, we can derive an upper bound as Inequality~\ref{eq:lasso-upper-bound}.
Furthermore, we can reduce the gap between the time cost and the upper bound by finding a tunnel $\left(c, W(s)\right)$ that minimizes the right-hand side of Inequality~\ref{eq:lasso-upper-bound}.
To minimize the right-hand side, we need to find a tunnel that is as short and narrow as possible.
We design an algorithm to estimate the tunnel that aims to minimize the upper bound estimation:

\begin{enumerate}[leftmargin=*]
    \item \textbf{Find inner contour:} Compute the convex hull of $X^+$, denoted as $c^+$.

    \item \textbf{Filter:} Remove all the points in $X^-$ enclosed by $c^+$.
          For a point $x^-$ not enclosed by $c^+$, the corresponding winding number is zero, i.e., $\mathrm{wind}(c^+, x^-) = 0$.

    \item \textbf{Interpolate:} Interpolate $c^+$ to get $c^+_{dense}$ with $N$ points where the points in $c^+_{dense}$ are evenly distributed.

    \item \textbf{Find outer contour:} For each $x_i^+ \in c^+_{dense}$, find the closest point in $X^-$ denoted as $x_i^-$.

    \item \textbf{Estimate medial axis:} For each $x_i^+ \in c^+_{dense}$, use $x_i = \frac{x_i^+ + x_i^-}{2}$ as a control point of the medial axis of the tunnel.

    \item \textbf{Estimate width:} For each $x_i^+ \in c^+_{dense}$, use the euclidean distance $d(x_i^+, x_i^-)$ as the tunnel width at $x_i$.
\end{enumerate}

Algorithm~\ref{alg:estimate-lasso-selection-tunnel} summarizes the pseudocode for estimating the lasso tunnel $(c, W(s))$.

\begin{algorithm}[htbp]
    \small
    \caption{EstimateLassoTunnel}\label{alg:estimate-lasso-selection-tunnel}
    \begin{algorithmic}[1]
        \Require positive and negative points $ X^+ $, $ X^- $
        \Ensure estimated lasso selection tunnel $ (c, W) $
        
        \State $ c^+ \gets \mathrm{convexHull}(X^+) $ \Comment{Find inner counter}
        \State $ X^- \gets \{x_i | x_i \in X^- \wedge \mathrm{wind}(c^+, x_i) = 0 \} $ \Comment{Filter}
        \State $ c^+ \gets \mathrm{interpolate}(c^+, N) $ \Comment{Interpolate}
        \State $ c^- \gets (\mathrm{argmin}_{x^-_i \in X^-} d(x^+_i, x^-_i))_{i=1}^N $ \Comment{Find outer contour}
        \State $ c \gets (\frac{x^+_i + x^-_i}{2})_{i=1}^N $ \Comment{Estimate medial axis}
        \State $ W \gets (d(x^+_i, x^-_i))_{i=1}^N $ \Comment{Estimate width}
        \State \Return{$ (c, W) $}
    \end{algorithmic}
\end{algorithm}

With the process above, we obtain a tunnel formed by a series of $N$ quadrilaterals.
The i-th quadrilateral has vertices $x_i^+$, $x_i^-$, $x_{i+1}^+$, and $x_{i+1}^-$.
(For notational convenience, throughout the writing, we let $x_{N+1} = x_1$ when the valid range of $i$ for $x_i$ is $\{1, 2, ..., N\}$.)
The time cost of passing the i-th quadrilateral can be estimated by a special case of general tunnel passing (Fig.~\ref{fig:bounds}(E)), the variable-width tunnel passing~\cite{Accot1997Fitts}, as $T_i = a_s + b_s ID_i$ where:

\begin{equation}\label{eq:id-variable-width-tunnel-passing}
    ID_i = \frac{D_i}{W_{i+1} - W_i} ln\frac{W_{i+1}}{W_i}
\end{equation}

\begin{itemize}[leftmargin=*]
    \item $D_i = d(x_i, x_{i+1})$ is the distance between the start and end lines.

    \item $W_i = d(x_i^+, x_i^-)$ is the width of the start line.

    \item $W_{i+1} = d(x_{i+1}^+, x_{i+1}^-)$ is the width of the end line.
\end{itemize}

Thus, given the estimated tunnel, we estimate the time cost of lasso selection by summing up the time cost of passing the $N$ quadrilaterals:

\begin{equation}\label{eq:t-lasso-estimation}
    \hat{T}_{lasso} = \sum_{i=1}^{N}(a_s + b_s ID_i)
\end{equation}

In Equation~\ref{eq:t-lasso-estimation}, $a_s$ and $b_s$ are constants with $b > 0$.
Thus, optimizing the estimated lasso selection time cost $\hat{T}_{lasso}$ is equivalent to optimizing $\hat{ID}_{lasso} = \sum_{i=1}^{N}ID_i$.

Note that the original variable-width tunnel passing task requires that the start and end lines are parallel~\cite{Accot1997Fitts}.
This condition approximately holds when the points are dense.
To satisfy this condition, we interpolate the points to increase the point density.

Algorithm~\ref{alg:estimate-lasso-selection-id} summarizes the pseudocode for estimating the index of difficulty of lasso selection $\hat{ID}_{lasso}$.

\begin{algorithm}[htbp]
    \small
    \caption{EstimateLassoID}\label{alg:estimate-lasso-selection-id}
    \begin{algorithmic}[1]
        \Require lasso selection tunnel $ c = (x_i)_{i=1}^N $, $ W = (W_i)_{i=1}^N $
        \Ensure estimated lasso selection index of difficulty $ \hat{ID}_{lasso} $
        
        \State $ D \gets (d(x_i, x_{x+1}))_{i=1}^N $
        \State $ ID \gets (\frac{D_i}{W_{i+1} - W_i} ln \frac{W_{i+1}}{W_i})_{i=1}^N $
        \State $ \hat{ID}_{lasso} \gets \sum_{i=1}^N ID_i $
        \State \Return{$ \hat{ID}_{lasso} $}
    \end{algorithmic}
\end{algorithm}

The lasso selection tunnel's inner contour may enclose \category{negative} points.
In such cases, the \category{positive} points are not separable from the \category{negative} points.
Let $N^-$ be the number of points in $X^-$ enclosed by $c^+$.
To penalize the viewpoint that encloses \category{negative} points, in the implementation, we multiply the index of difficulty with $exp(\frac{20 N^-}{|c^+|})$.

\subsubsection{Time Complexity Analysis}
\label{sec:lasso-cost-estimation-time-complexity}

Let $n$ be the number of points visible under a given viewpoint.
The time complexity of computing the convex hull to find the inner contour is $O(nlogn)$~\cite{Sklansky1982Finding}.
The time complexity to find the outer contour is $O(n^2)$.
The time complexities of interpolation, medial axis estimation, width estimation, and computing the time cost to pass the quadrilaterals are all $O(N)$.
Typically, we set $N$ to be much smaller than $n^2$.
Thus, the time complexity of the lasso time cost estimation is $O(n^2)$.
The bottleneck is finding the outer contour.

\begin{figure}[!htbp]
    \centering
    \includegraphics[width=0.8\linewidth]{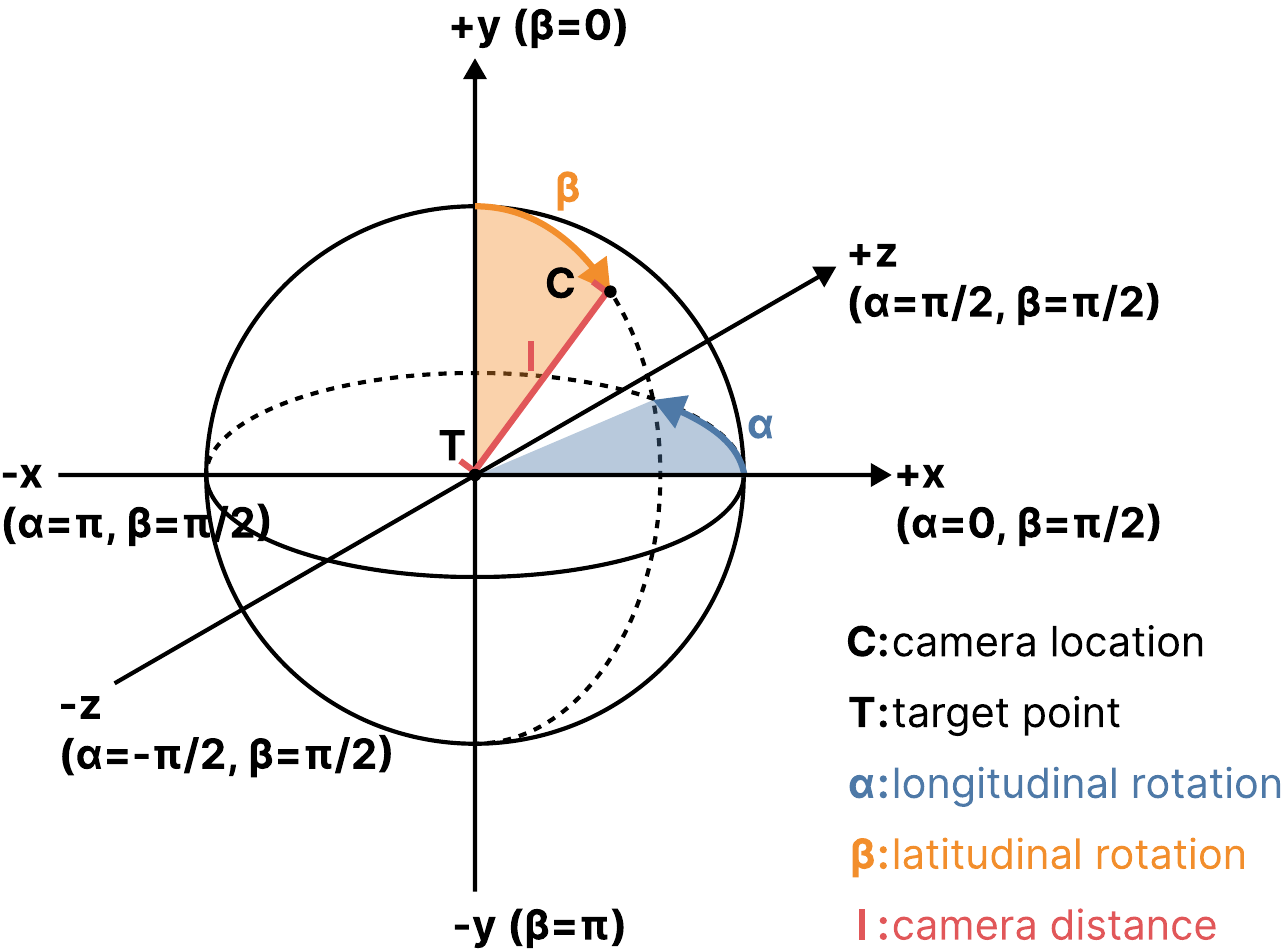}
    \caption{
        \textbf{The parameters of an arc rotate camera: $\mathbf{\alpha}$, $\mathbf{\beta}$, and $\mathbf{l}$.}
        The camera is placed at point $C$ and targets at point $T$.
        $\alpha$ is the longitudinal rotation of the camera.
        $\beta$ is the latitudinal rotation of the camera.
        $l$ is The distance between the camera and the target.
    }
    \label{fig:arc-rotate-camera}
\end{figure}

\begin{figure*}
    \centering
    \includegraphics[width=\linewidth]{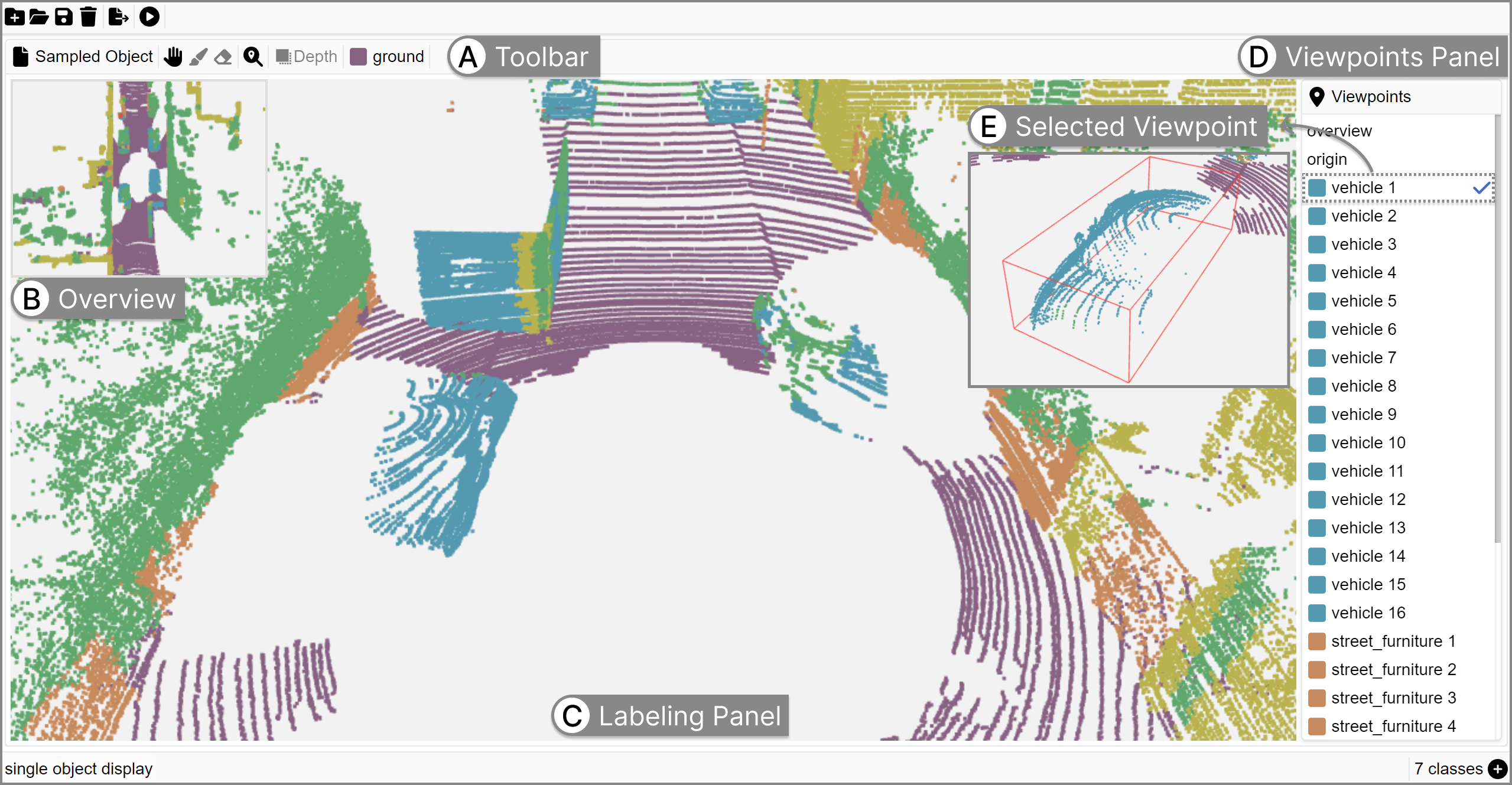}
    \caption{
        \textbf{The interface of our point cloud labeling system:}
        (A) The toolbar contains widgets for switching among interaction modes (navigate, label, and erase) and switching the label category to assign to lasso selected points.
        (B) The overview shows the point cloud from a top-down viewpoint.
        (C) The labeling panel is the workspace where the user moves the camera and annotates the point cloud.
        (D) The viewpoints panel shows recommended viewpoints for each object.
        (E) Selecting a recommended viewpoint moves the camera of the labeling panel.
    }
    \label{fig:interface}
\end{figure*}

\subsection{Grid Searching Optimal Viewpoint}
\label{sec:viewpoint-optimization}

Using the lasso selection time cost model in Sec.~\ref{sec:lasso-time-cost}, we evaluate the quality of a viewpoint, given the instance to label.
Thus, we optimize the viewpoint with a grid search of the viewpoint parameters on the viewing sphere that maximize the quality.

For a viewpoint represented as an arc rotate camera as shown in Fig.~\ref{fig:arc-rotate-camera}, its parameters are:

\begin{itemize}[leftmargin=*]
    \item \textbf{Target point} $\mathbf{T}$:
          The target point in the 3D space with coordinates $x_t \in \mathbb{R}^3$ that the camera points at.

    \item \textbf{Longitudinal rotation} $\bm{\alpha}$:
          The camera rotation in the longitudinal axis.
          Its value range is $(-\pi, \pi]$ (in radians).

    \item \textbf{Latitudinal rotation} $\bm{\beta}$:
          The camera rotation in the latitudinal axis.
          Its value range is $[0, \pi]$ (in radians).

    \item \textbf{Camera distance} $\mathbf{l}$:
          The distance in the 3D space between the camera location $C$ and the target point $T$.
          Its value range is $\mathbb{R}^+$.
\end{itemize}

We choose the gravity center of an object as the target point $T$.
Given a viewpoint, the 3D points can be projected to 2D for display.

\subsubsection{Optimizing Camera Rotation}
\label{sec:optimizing-camera-rotation}

For longitudinal rotation $\alpha$, we grid search in $(-\pi, \pi]$ with a stride of $\pi/12$.
For latitudinal rotation $\beta$, we grid search in $[0, \pi]$ with a stride of $\pi/12$.
One may use a smaller stride in the grid search to achieve a more accurate result if needed.
In the grid search, we use the estimated lasso selection time cost $\hat{T}_{lasso}$ as the loss function to be minimized, which is equivalent to minimizing $\sum_{i=1}^{N}ID_i$ according to Sec.~\ref{sec:estimation-algorithm}:

\begin{equation}
    (\alpha, \beta)
    = \underset{(\alpha, \beta)}{argmin} \ \sum_{i=1}^{N}ID_i(\alpha, \beta)
\end{equation}

Algorithm~\ref{alg:optimize-viewpoint} summarizes the pseudocode for grid searching the optimal viewpoint $(\alpha, \beta)$.

\begin{algorithm}[htbp]
    \small
    \caption{OptimizeViewpoint}\label{alg:optimize-viewpoint}
    \begin{algorithmic}[1]
        \Require $ P_3 = \{(x_i, y_i)\}_{i=1}^n $ where $x_i \in \mathbb{R}^3 \wedge y_i \in \{1, -1\}$
        \Ensure optimized camera rotation $ (\alpha^*, \beta^*) $
        
        \State $ P^+_3 \gets \{(x_i, y_i) | (x_i, y_i) \in P_3 \wedge y_i = 1\} $
        \State $ x_t \gets \frac{1}{|P^+_3|} \sum_{(x_i, y_i) \in P^+_3} x_i $ \Comment{Gravity center of target object}
        \State $ (\alpha^*, \beta^*, ID^*) \gets (\mathrm{null}, \mathrm{null}, +\infty) $ \Comment{Record of best rotation}
        \For{$ \alpha \in (-\frac{11\pi}{12}, -\frac{10\pi}{12}, ..., \pi) $} \Comment{Grid search $\alpha$}
            \For{$ \beta \in (0, \frac{\pi}{12}, ..., \pi) $} \Comment{Grid search $\beta$}
                \State $ P \gets \{((M(x_i - x_t))_{(1, 2)}, y_i)\}_{i=1}^n $ \Comment{Project to 2D}
                \State $ X^+ \gets \{x_i | (x_i, y_i) \in P \wedge y_i = 1\} $
                \State $ X^- \gets \{x_i | (x_i, y_i) \in P \wedge y_i = -1\} $
                \State $ (c, W) \gets \mathrm{EstimateLassoTunnel}(X^+, X^-) $
                \State $ ID \gets \mathrm{EstimateLassoID}(c, W) $
                \If{$ ID < ID^* $}
                    \State $(\alpha^*, \beta^*, ID^*) \gets (\alpha, \beta, ID)$
                \EndIf
            \EndFor
        \EndFor
        \State \Return{$ (\alpha^*, \beta^*) $}
    \end{algorithmic}
\end{algorithm}

\subsubsection{Choosing Camera Distance}

Fitts' law time cost models for mouse movement tasks are invariant to scaling, as can be seen from the equations in Sec.~\ref{sec:fitts-law}.
As the camera moves closer to an object, both $W(s)$ and $c$ in Inequality~\ref{eq:lasso-upper-bound} scale with the same factor.
As Fitts' law does not suggest a specific camera distance $l$, we compute $l$ with the following heuristics:

Typically, an annotator may move the camera closer when annotating smaller objects.
Under this observation, we assume that annotators would prefer the annotated object to occupy a fixed portion of the field of view, regardless of the actual size of the annotated object and the point density.

Thus, we set the camera distance as $1.5d_{diag}$.
$d_{diag}$ is the length of the diagonal of the axes-aligned bounding cuboid of the object.

\subsubsection{Time Complexity Analysis}

The bottleneck of the grid search is the lasso selection time cost estimation.
As discussed in Sec.~\ref{sec:lasso-cost-estimation-time-complexity}, the time complexity of lasso selection time cost estimation is $O(n^2)$.
Let $q$ be the number of combinations of $(\alpha, \beta)$ evaluated in the grid search.
The time complexity of the grid searching process is $O(q n^2)$.
The recommended viewpoints can be precomputed before the user starts labeling.
Thus, the time complexity is typically not a concern in practice.
If necessary, one may use alternative optimization approaches, such as gradient-based optimization, to reduce the time complexity.

\subsection{Assumptions and Considerations}
\label{sec:assumptions-considerations}

In the following, we discuss the assumptions and considerations in our time cost modeling.

\textbf{Requirement on initial segmentation:}
As Sec.~\ref{sec:estimation-algorithm} introduces, our lasso time cost estimation algorithm requires that each point has an initial segmentation label.
We should not expect the initial segmentation to be perfect.
Otherwise, there would be no need for labeling.
Meanwhile, the initial segmentation should be good enough to roughly group points into semantic instances.
Otherwise, our time cost modeling and the corresponding viewpoint recommendation may not function as expected.
In other words, our approach is limited to the scenario of label refinement and cannot handle the scenario of labeling from scratch.
When an instance segmentation model is used to provide default labels, the objects are readily grouped.
When a semantic segmentation model is used to provide default labels, we group points with DBSCAN~\cite{Ester1996Density}, as discussed in
\ifx\hideappendix\undefined
    Appendix~\ref{sec:point-cloud-clustering}.
\else
    Appendix B in the supplementary materials.
\fi

\textbf{Requirement on object size:}
Our lasso time cost model assumes that the points to be selected and the other points form two separable classes.
This assumption likely fails for large objects, such as \category{ground} in a point cloud of a street scene.
We recommend not to use our approach for the objects that are expected to be large or expected to mix with points of other categories.

\textbf{Auto-closing function:}
The lasso selection interaction is often implemented with an auto-closing function.
The auto-closing function closes the lasso selection when the user releases the mouse.
Auto-closing may reduce the time cost of lasso selection.
Our time cost modeling in Section~\ref{sec:lasso-time-cost} ignores the impact of auto-closing on the time cost.

\textbf{Alternative time cost model:}
Before our work, Yamanaka et al.~\cite{Yamanaka2022Effectiveness,Yamanaka2019Modeling} have also worked on modeling the time cost of lasso selection.
They focus on scenarios such as lasso selection in an aligned matrix of rectangular icons and assume the tunnel is known in advance.
Thus, their model does not apply to lasso selection in point clouds.

\textbf{Interpretation from class separability:}
As Sec.~\ref{sec:estimation-algorithm} introduces, we aim to find a viewpoint that minimizes the estimated upper bound of lasso selection time cost.
As Sec.~\ref{sec:lasso-time-cost} introduces, finding such a viewpoint, roughly speaking, is finding a viewpoint that minimizes $\frac{D}{W}$ where $D$ is the length of the lasso selection tunnel, and $W$ is the width of the tunnel.
From the classification perspective, $W$ is the margin between the \category{positive} and \category{negative} categories.
In other words, our approach favors viewpoints with good separability between the categories.
In this sense, our optimization goal is related to the goal of the support vector machine, which is to minimize $\frac{1}{W}$.

  \begin{figure*}[htbp]
    \centering
    \includegraphics[width=\linewidth]{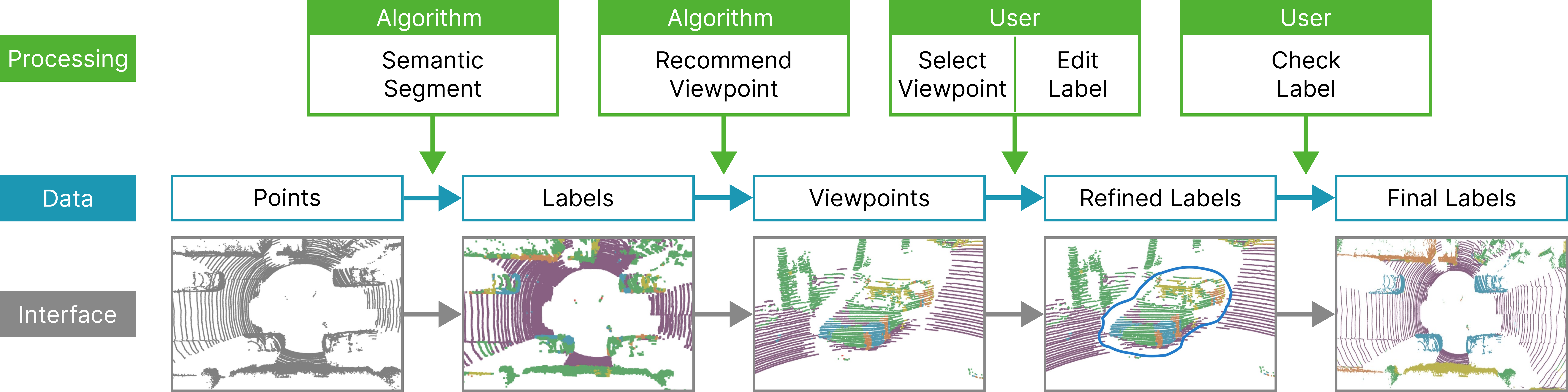}
    \caption{
        \textbf{System workflow:}
        With the input 3D point cloud, the system first generates pre-labeled results using a segmentation model and recommended viewpoint for each semantic instance.
        The annotator selects the view from the recommended viewpoints and uses lasso selection to refine the annotation.
        The refined labels are obtained after reviewing the refined labels.
    }
    \label{fig:workflow}
\end{figure*}

\section{Point Cloud Labeling System}
\label{sec:data-labeling-system}

We develop a point cloud labeling system that integrates our viewpoint recommendation approach, as shown in Fig.~\ref{fig:interface}.
Figure~\ref{fig:workflow} shows the system workflow.
The system provides a top-down overview of the point cloud (Fig.~\ref{fig:interface}(B)).
For the overview, the camera is placed at a distance of $1.5d'_{diag}$ from the origin.
$d'_{diag}$ is the length of the diagonal of the axes-aligned bounding cuboid of the whole point cloud.
The user can modify the point labels through lasso selection in the labeling panel (Fig.~\ref{fig:interface}(C)).
The user can pan, zoom, and rotate to move the camera and change the viewpoint.
The system is developed with the OneLabeler~\cite{Zhang2022OneLabeler} framework.

\textbf{Default labeling:}
To reduce the burden of labeling, the system assigns default labels to the points before the annotator starts labeling.
The default labels for semantic segmentation are predicted with PolarNet~\cite{Zhang2020PolarNet}.
If needed, the user may switch to other models, such as PointNet~\cite{Charles2017PointNet}.

\textbf{Viewpoint recommendation:}
The system provides a list of recommended viewpoints for the annotator to choose in the viewpoints panel (Fig.~\ref{fig:interface}(D)).
Each recommended viewpoint corresponds to a semantic object, e.g., \category{vehicle}.
Selecting a recommended viewpoint in the list moves the camera of the labeling panel to the viewpoint through a smooth animation.
The corresponding object of the recommended viewpoint is highlighted with a bounding cuboid.
When users modify labels in this view, the corresponding box on the right will be ticked, denoting that the viewpoint has been checked.

  \section{Ablation Study}
\label{sec:ablation-study}

We carried out an ablation study to examine the effectiveness of our viewpoint recommendation approach.
The study compared the time cost and label quality when using different viewpoint recommendation approaches.
We also gathered subjective feedback from the participants.

\begin{figure*}
    \centering
    \includegraphics[width=\linewidth]{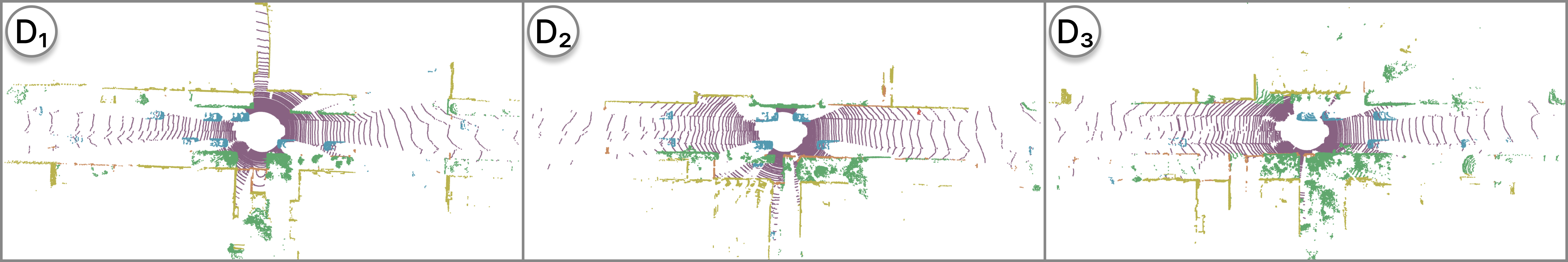}
    \caption{
        Three point clouds from the \dataset{SemanticKITTI} dataset were used in the study (denoted as \dataset{D1}, \dataset{D2}, and \dataset{D3}).
    }
    \label{fig:user-study-point-clouds}
\end{figure*}

\subsection{Experiment Design}

Our approach in Sec.~\ref{sec:approach} can be decomposed into two parts: target object recommendation and camera rotation recommendation.
In the ablation study, we aim to examine the effectiveness of the two parts.

\textbf{Participants:}
We recruited nine participants for the study, denoted as P1, P2, ..., P9.
All of them were students or recent graduates majoring in data science or computer science.
They all had experience using 3D graphics interfaces, such as 3D modeling software.
Five (P2, P3, P5, P7, P9) had research experience labeling or processing point cloud data.

\textbf{Compared methods:}
We compare three methods.
For each method, the viewpoint is initialized at the origin of the coordinate system and follows the negative direction of the x-axis.

\begin{itemize}[leftmargin=*]
    \item \textit{No Recommendation}:
          The labeling system provides no viewpoint recommendation.

    \item \textit{Target}:
          The labeling system recommends a list of viewpoints.
          Each recommended viewpoint targets the center of a semantic instance with a random camera rotation.

    \item \textit{Target + Rotation}:
          The labeling system recommends a list of viewpoints.
          Each recommended viewpoint targets the center of a semantic instance with the camera rotation computed with the viewpoint recommendation approach.
          This method is the one proposed in Sec.~\ref{sec:approach}.
\end{itemize}

\textbf{Metrics:}
We compare the three methods with the completion time, the increase of the label's mean intersection over union (delta mIoU), and the user score as the metrics.
We use mIoU instead of accuracy to quantify the label quality, as the distribution of points in categories can be highly imbalanced.
Some categories (e.g., \category{vehicle}) may take a much smaller share than others (e.g., \category{ground}).

\textbf{Hypotheses:}
We set the following hypotheses to be examined in the study.

\begin{itemize}[leftmargin=*]
    \item \textit{H1:}
          Target recommendation and rotation recommendation both reduce completion time.
          ($t_{no} > t_{target} > t_{target + rotation}$)

    \item \textit{H2:}
          The recommendation method does not influence label quality as quantified by delta mIoU.
          ($dm_{no} \approx dm_{target} \approx dm_{target + rotation}$)

    \item \textit{H3:}
          Target recommendation and rotation recommendation both improve user satisfaction.
          ($s_{no} < s_{target} < s_{target + rotation}$)
\end{itemize}

\textbf{Apparatus:}
The experiment was conducted within the data labeling interface introduced in Sec.~\ref{sec:data-labeling-system} on an HD display (1920px $\times$ 1080px).
The participants used a standard mouse with 800 DPI as the input device.

\textbf{Dataset:}
The study is conducted on the \dataset{SemanticKITTI} dataset.
The used point clouds are downsampled to 100k points.
\dataset{SemanticKITTI} has six major point categories: \category{ground}, \category{structure}, \category{vehicle}, \category{nature}, \category{human}, and \category{object}.
To make the study session concise, we focus on the semantic segmentation of the point cloud into two classes: \category{vehicle} and \category{not\_vehicle}.
We randomly selected three point clouds from the training set of \dataset{SemanticKITTI} (referred to as \dataset{D1}, \dataset{D2}, \dataset{D3}), as shown in Fig.~\ref{fig:user-study-point-clouds}.
We used PolarNet~\cite{Zhang2020PolarNet} with dropout being 0.5 to assign the default binary semantic segmentation labels to the points.
The mIoU of the default labeling results for \dataset{D1}, \dataset{D2}, and \dataset{D3} is $0.6756$, $0.5680$, and $0.6432$.

\textbf{Procedure:}
For each participant, the study session lasts for about 45 minutes.

\begin{enumerate}[leftmargin=*]
    \item \textbf{Pre-study questionnaire:}
          We ask the participant about the experience with 3D graphics interfaces and data labeling ($\sim$5 minutes).

    \item \textbf{Training:}
          We train the participant to use our labeling system ($\sim$15 minutes).
          In training, we first play a video demonstration and show a corresponding system manual to teach the participant the major functions of the labeling system, such as moving the camera, selecting points with lasso selection, and choosing recommended viewpoints.
          Then, we ask the participant to practice using the system to label a point cloud from \dataset{SemanticKITTI}.
          After finishing labeling, we show the participant the ground truth label and ask the participant to articulate their mistakes to improve their labeling skills.

    \item \textbf{Task:}
          We instruct the participant to label three point clouds from \dataset{SemanticKITTI} with three methods ($\sim$15 minutes).
          The participant is instructed to segment each point cloud into \category{vehicle} and \category{not\_vehicle}.
          Figure~\ref{fig:user-study-point-clouds} shows the three point clouds.
          For each point cloud, a different viewpoint recommendation approach (\userStudyMethod{no recommendation}, \userStudyMethod{target}, and \userStudyMethod{target + rotation}) is used.
          To offset the learning effect, we randomize the order of the three point clouds and three methods for different participants.
          The participant is instructed to annotate as fast and accurately as possible and focus more on speed.
          The participant is advised to utilize the recommended viewpoints when recommendations are available.
          The completion time and delta mIoU are recorded.
          We start recording the time when the participant begins to label and stop when the participant reports completion.

    \item \textbf{Post-study questionnaire:}
          We ask the participant to score the labeling experience with a 7-point Likert scale for each point cloud and interview the participant ($\sim$10 minutes).
\end{enumerate}

\begin{figure}[t]
    \centering
    \includegraphics[width=\linewidth]{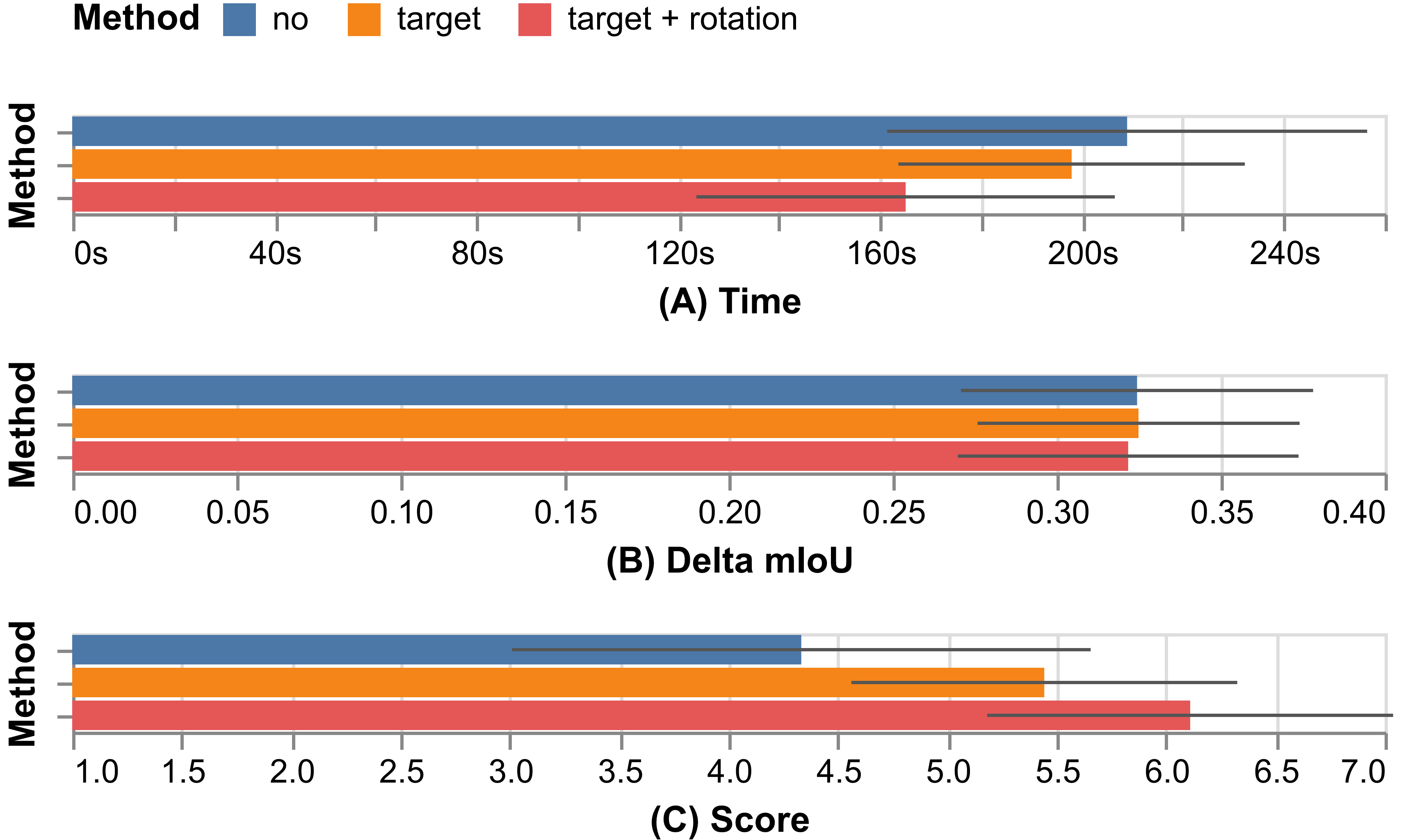}
    \caption{
        \textbf{The time, delta mIoU, and score of different methods:}
        \userStudyMethod{Target + rotation} exhibits reduced completion time and increased user preference score, compared with \userStudyMethod{target} and \userStudyMethod{no recommendation}.
        The error bar shows the standard deviation.
    }
    \label{fig:experiment}
\end{figure}

\subsection{Quantitative Results}

For the time, delta mIoU, and score averaged on the three point clouds, we use ANOVA to examine whether there are significant differences between \userStudyMethod{no recommendation}, \userStudyMethod{target}, and \userStudyMethod{target + rotation}.

\textbf{Time:}
Figure~\ref{fig:experiment}(A) shows the average time of the three methods.
$t_{no} = 209.0 \pm 47.5$s, $t_{target} = 198.0 \pm 34.3$s, and $t_{target + rotation} = 165.1 \pm 41.4$s.
We observe no significant difference between $t_{no}$ and $t_{target}$ ($p = 5.81 \times 10^{-1}$).
Asides, we obverse $t_{target} > t_{target + rotation}$ ($p = 8.54 \times 10^{-2}$) and $t_{no} > t_{target + rotation}$ ($p = 5.32 \times 10^{-2}$).
Hypothesis \textit{H1} is partially validated.

\textbf{Delta mIoU:}
Figure~\ref{fig:experiment}(B) shows the average delta mIoU of the three methods.
$dm_{no} = 0.3246 \pm 0.0537$, $dm_{target} = 0.3250 \pm 0.0491$, and $dm_{target + rotation} = 0.3218 \pm 0.0519$.
We observe no significant differences between $dm_{no}$ and $dm_{target}$ ($p = 9.85 \times 10^{-1}$), $dm_{no}$ and $dm_{target + rotation}$ ($p = 9.15 \times 10^{-1}$), and $dm_{target}$ and $dm_{target + rotation}$ ($p = 8.96 \times 10^{-1}$).
Hypothesis \textit{H2} is validated.

\textbf{Score:}
Figure~\ref{fig:experiment}(C) shows the average score of the three methods.
$s_{no} = 4.33 \pm 1.32$, $s_{target} = 5.44 \pm 0.88$, and $s_{target + rotation} = 6.11 \pm 0.93$.
We observe $s_{no} < s_{target}$ ($p = 5.23 \times 10^{-2}$) and $s_{no} < s_{target + rotation}$ ($p = 4.51 \times 10^{-3}$).
Meanwhile, there is no significant difference between $s_{target}$ and $s_{target + rotation}$ ($p = 1.38 \times 10^{-1}$).
In summary, target recommendation improves user satisfaction, while rotation recommendation does not significantly influence user satisfaction.
Hypothesis \textit{H3} is partially validated.

\subsection{Participant Feedback}

All the participants commented that the viewpoint recommendation was helpful.
Specifically, P1, P4, and P7 noted that these recommendations reduced the time needed to adjust the camera to locate vehicles compared to manually searching.
Meanwhile, P4 mentioned that the recommendations missed some vehicles, which may cause labeling mistakes due to the tendency to rely on the recommendations.

The participants also suggested potential usability improvements.
P2 and P5 suggested that the system show more information about the recommended viewpoints so that they may focus on potential high-quality recommendations.
P3, P6, and P9 suggested adding keyboard shortcuts for switching between the navigation and labeling modes.

  \begin{figure*}[t]
    \centering
    \includegraphics[width=\linewidth]{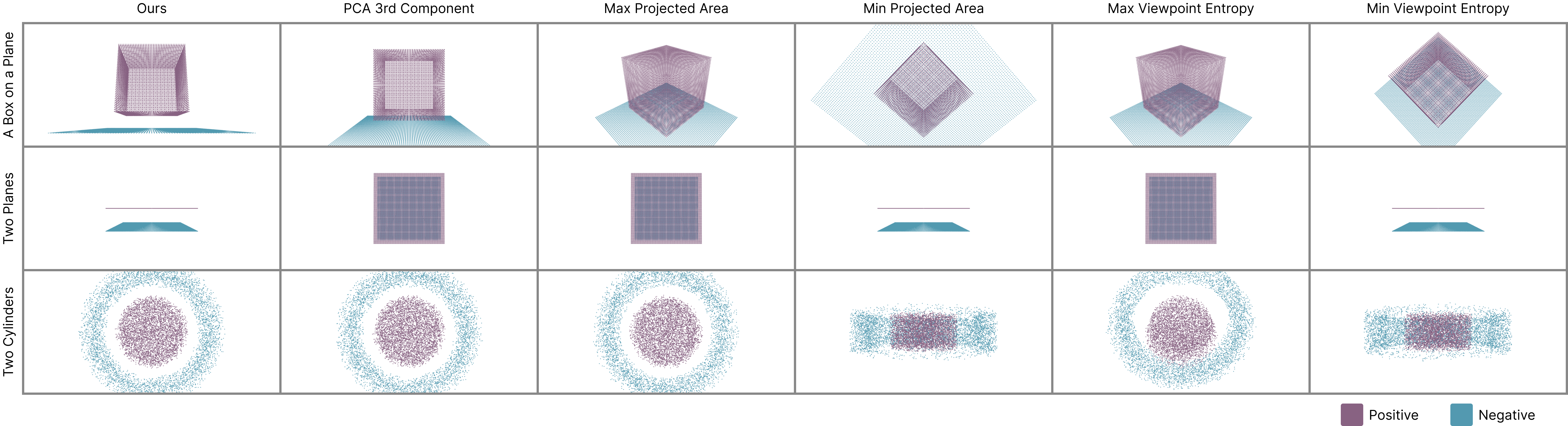}
    \caption{
        \textbf{Recommended viewpoints for synthetic data:}
        The recommended viewpoints for the \category{positive} category in three synthetic point clouds: \dataset{a box on a plane}, \dataset{two planes}, and \dataset{two cylinders}.
        For the three point clouds, our method selects viewpoints that do not overlap the \category{positive} and \category{negative} categories.
        For \dataset{two planes}, the min viewpoint entropy and min projected area recommend viewpoints that effectively separate the two categories.
        Meanwhile, they recommend viewpoints that overlap the two categories for \dataset{two cylinders}.
    }
    \label{fig:comparison-synthetic}
\end{figure*}

\section{Comparative Analysis}
\label{sec:comparative-analysis}

This section demonstrates a qualitative comparison between our method and other viewpoint selection methods.
To highlight the characteristics of different viewpoint selection methods, we assume the true semantic segmentation labels are known.
We conduct the comparison on synthetic datasets as well as indoor and outdoor real-world datasets.
Specifically, we compare our method with PCA, projected area~\cite{Secord2011Perceptual}, and viewpoint entropy~\cite{Vazquez2001Viewpoint}.

\begin{itemize}[leftmargin=*]
    \item \textbf{PCA:}
          The third principal component is commonly used as a viewpoint to normalize the pose of 3D objects~\cite{Kim2013new,Li2021Closer}.
          It maximizes the spatial variance of the points.
          The following refers to this method as PCA 3rd component.

    \item \textbf{Projected area:}
          This method selects the viewpoint that maximizes the projected area of 3D objects~\cite{Secord2011Perceptual}.
          The following refers to this method as max projected area.
          We also consider an opposite method that minimizes the projected area.
          The following refers to the opposite method as min projected area.

    \item \textbf{Viewpoint entropy:}
          This method selects the viewpoint that maximizes the entropy of projected surfaces~\cite{Vazquez2001Viewpoint}.
          The following refers to this method as max viewpoint entropy.
          We also consider an opposite method that minimizes the viewpoint entropy.
          The following refers to the opposite method as min viewpoint entropy.
\end{itemize}

Viewpoint entropy and projected area are designed for mesh models.
Before using them, we use Katz et al.'s method to remove hidden points~\cite{Katz2007Direct} and convert each semantic instance into a mesh model.

\subsection{Comparison on Synthetic Datasets}

We used three synthetic point clouds for comparison, as shown in Fig.~\ref{fig:comparison-synthetic}.
In each point cloud, the points belong to two label categories: \category{positive} and \category{negative}.
Each category consists of one semantic instance comprising all the points in this category.
We consider recommending viewpoints for the semantic instance of the \category{positive} category.
The following describes the setup of the three synthetic point clouds.

\begin{itemize}[leftmargin=*]
    \item \dataset{A box on a plane}:
          The \category{positive} points form a cuboid surface.
          The \category{negative} points form a plane.
          The plane parallels the bottom of the cuboid.
          The points in the cuboid surface and plane are uniformly spaced.

    \item \dataset{Two planes}:
          The \category{positive} points form a plane.
          The \category{negative} points form another plane.
          The two planes are parallel.
          The points in each plane are uniformly spaced.

    \item \dataset{Two cylinders}:
          The \category{positive} points form an inner solid cylinder.
          The \category{negative} points form an outer hollow cylinder.
          The two planes are concentric.
          The points in each plane are uniformly distributed.
\end{itemize}

As shown in Fig.~\ref{fig:comparison-synthetic}, for all the three point clouds, our method selects viewpoints that leave a large margin between the \category{positive} and \category{negative} points.
As Fitts' law implies, a larger margin reduces the difficulty of the mouse movement.
By comparison, the other methods may select viewpoints that overlap the two categories.
In point cloud labeling, the overlap between categories may make it hard for annotators to select the \category{positive} points.

Min projected area and min viewpoint entropy favor viewpoints that overlap the point features.
For \dataset{two planes}, our method, min projected area, and min viewpoint entropy all select a viewpoint that degenerates the plane of \category{positive} points into a line.
PCA 3rd component, max projected area, and max viewpoint entropy favor viewpoints that provide more information about the point features.
For \dataset{two cylinders}, our method follows PCA 3rd component and max projected area and selects a viewpoint along the central axis.

\begin{figure}[hb]
    \centering
    \includegraphics[width=\linewidth]{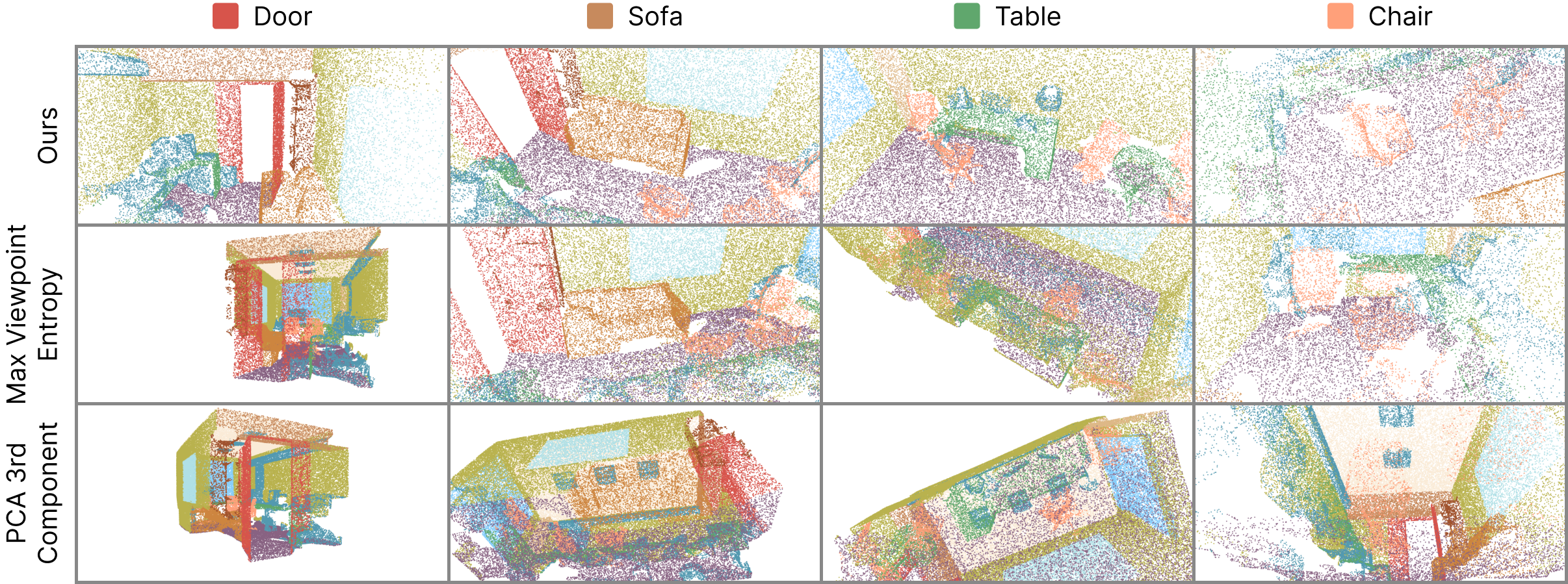}
    \caption{
        \textbf{Recommended viewpoints for \dataset{S3DIS}:}
        The recommended viewpoints for five categories: \category{door}, \category{sofa}, \category{table}, \category{chair}, and \category{wall}.
        Compared with max viewpoint entropy and PCA 3rd component, our method tends to select viewpoints with less occlusion.
    }
    \label{fig:comparison-S3DIS}
\end{figure}

\subsection{Comparison on Real-World Datasets}

We made the comparison on \dataset{Semantic3D}~\cite{Hackel2017SEMANTIC3D.net} and \dataset{S3DIS}~\cite{Armeni20163D}.
Each point cloud was downsampled to 100k points prior to the viewpoint recommendation.

\textbf{Viewpoint Recommendation for \dataset{S3DIS}:}
Figure~\ref{fig:comparison-S3DIS} shows the recommended viewpoints for objects in the \dataset{S3DIS} indoor scene dataset~\cite{Armeni20163D}.
The points of indoor objects are enclosed by \category{wall} for indoor scenes.
Thus, the points can be heavily occluded.
For indoor objects, such as \category{table}, \category{chair}, and \category{wall}, our method selects viewpoints with less occlusion than max viewpoint entropy and PCA 3rd component.
For thin objects such as \category{door} and \category{wall}, our method tends to recommend a top-down viewpoint where the object size is minimized.
In this way, the object size appears minimized, which reduces overlap and shortens the mouse movement path for lasso selection.

\begin{figure}[ht]
    \centering
    \includegraphics[width=\linewidth]{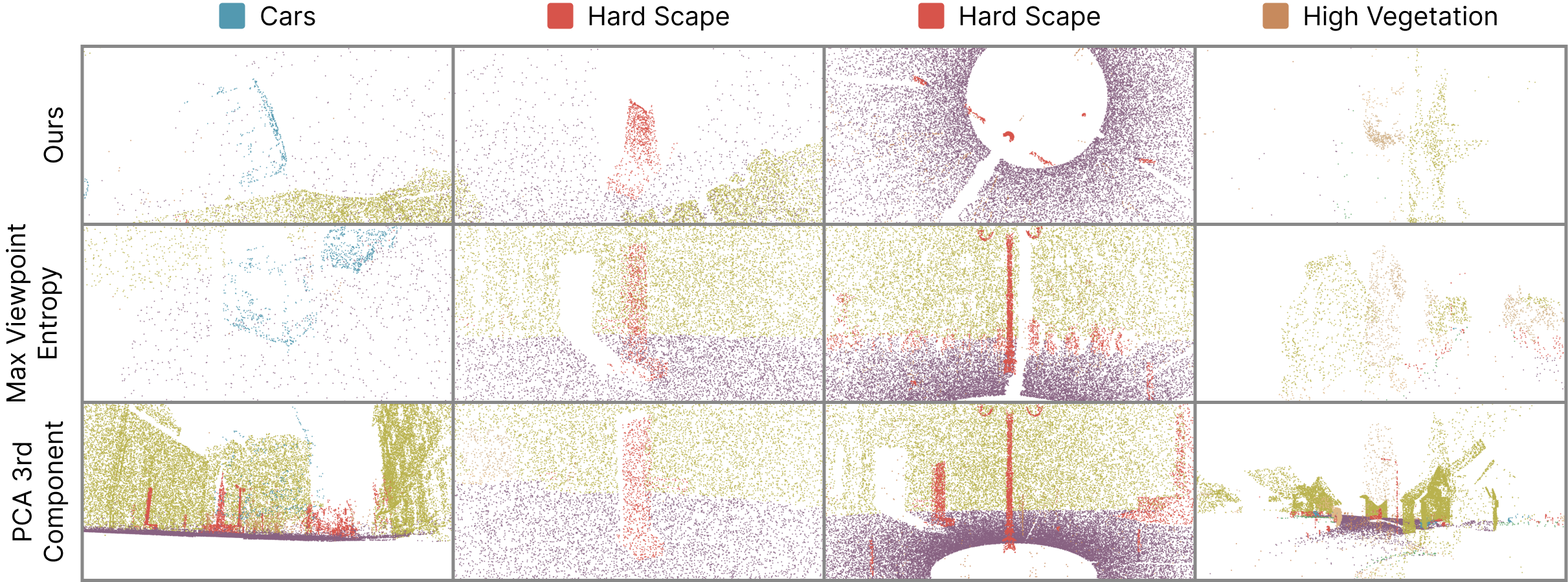}
    \caption{
        \textbf{Recommended viewpoints for \dataset{Semantic3D}:}
        The recommended viewpoints for three categories: \category{cars}, \category{hard scape}, and \category{high vegetation}.
        For thin objects belonging to \category{hard scape} and \category{high vegetation}, our method tends to recommend a top-down viewpoint.
    }
    \label{fig:comparison-Semantic3D}
\end{figure}

\textbf{Viewpoint Recommendation for \dataset{Semantic3D}:}
Figure~\ref{fig:comparison-Semantic3D} shows the recommended viewpoints for objects in the \dataset{Semantic3D} outdoor scene dataset~\cite{Hackel2017SEMANTIC3D.net}.
Compared with max viewpoint entropy and PCA 3rd component, our method tends to select viewpoints with less occlusion.
Our method tends to select a top-down viewpoint for thin objects, such as \category{hard scape} and \category{high vegetation}.

  \section{Discussion}
\label{sec:discussion}

In the following, we discuss the limitations of our work and potential future work.

\textbf{Improving the lasso selection time cost model:}
In Sec.~\ref{sec:approach}, we propose a lasso selection time cost model based on Fitts' law.
This model may be improved in the following aspects:

\begin{itemize}[leftmargin=*]
    \item \textbf{Considering error tolerance:}
          Our approach assumes \textit{error free} lasso selection, i.e., the lasso selection selects only points of one label category.
          In practice, a lasso selection may not be error-free.
          In some cases, the annotator may first conduct a selection with errors and then use subsequent selections to correct errors.
          In other cases without strict requirements on the label quality, the errors may be acceptable.

    \item \textbf{Considering multi-step lasso selection:}
          This work models the time cost of a single lasso selection.
          Meanwhile, multiple lasso selection is inevitable for semantic instances with holes, typically the background objects (e.g., \category{ground}).
          Future efforts may investigate the time cost of multi-step lasso selections.

    \item \textbf{Labeling multiple objects:}
          Our approach optimizes the time cost to label one object.
          An alternative setup is to recommend a viewpoint that jointly minimizes the time cost to label multiple objects.
\end{itemize}

\textbf{Validating the lasso selection time cost model:}
Our time cost model in Sec.~\ref{sec:approach} is derived from Fitts' law~\cite{Fitts1992Information} and steering law~\cite{Accot1997Fitts}, which are empirical laws.
Future user experiments may examine the validity of our model, like the experiment for establishing the steering law~\cite{Accot1997Fitts}.

\textbf{Optimizing the time costs of mental operators:}
As shown in Fig.~\ref{fig:lasso-labeling-steps}, the point cloud labeling involves not only physical operators for taking actions (e.g., \textit{adjust camera viewpoint} and \textit{draw lasso polygon(s)}) but also mental operators for making judgments (e.g., \textit{identify object(s)}).
This work focuses on optimizing the time cost of physical operators while the time cost of mental operators is not considered.
Future work may consider jointly optimizing the time costs of both physical and mental operators.
For example, the viewpoint recommender may select a viewpoint that is not only easy for the annotator to draw lasso polygons (i.e., optimize the time cost of physical operators) but also easy for the annotator to recognize objects (i.e., optimize the time cost of mental operators).
To model the time cost for annotators to recognize objects, previous work on canonical views of objects~\cite{Blanz1999What,Konkle2011Canonical} may be utilized.

\textbf{Connection with model-based evaluation:}
Our time cost modeling falls in the category of model-based evaluation~\cite{Sears2007Human} that focuses on using models to estimate usability metrics.
Early examples of model-based evaluation methods include the Keystroke-Level Model~\cite{Card1980Keystroke}.
Our work extends Zhang et al.'s model-based evaluation approach for quality-assurance interface~\cite{Zhang2019Cost,Zhang2021MI3,Zhang2024Simulation} to point cloud labeling.

\textbf{Visualization for routine tasks:}
Many efforts in visualization research are spent on investigating systems and techniques for analyzing data and communicating insights.
Meanwhile, using visualization for routine tasks, such as data labeling in this work, has received less attention.
There are various opportunities for visualization for routine tasks.
A particular opportunity is that the routine tasks are well-defined.
It is thus feasible to theoretically model user interaction and optimize the design for routine tasks, as discussed in this work and previous work~\cite{Zhang2019Cost,Zhang2021MI3,Zhang2024Simulation}.

  \section{Conclusion}

In this paper, we propose a viewpoint recommendation approach to reduce the time cost of point cloud labeling.
We use Fitts' law to model the time cost of lasso selection in 2D scatter plots.
Using the time cost model, we recommend viewpoints by grid searching for the viewpoints that minimize the lasso selection time cost.
We developed a point cloud labeling system that integrates our viewpoint recommendation approach.
An ablation study suggests that our approach effectively reduces the labeling time cost.
We also qualitatively compare our approach with other viewpoint recommendation approaches.
Our work applies model-based evaluation to the design of a data labeling system.
We envision various research opportunities for applying model-based evaluation to the design of interactive systems for routine tasks.

  \section*{Supplemental Material}

\ifx\hideappendix\undefined
    Appendix~\ref{sec:proofs}.
\else
    Appendix A, titled ``Proofs'',
\fi
describes the proofs of the mathematical derivations.
\ifx\hideappendix\undefined
    Appendix~\ref{sec:point-cloud-clustering}.
\else
    Appendix B, titled ``Point Cloud Clustering'',
\fi
describes our strategy to derive instance segmentation when a semantic segmentation model is used to provide default labels.

  \ifCLASSOPTIONcompsoc
  \section*{Acknowledgments}
\else
  \section*{Acknowledgment}
\fi

The authors thank Yang~Zhang, Xuan~Chen, Yuhan~Guo, Xinghua~Jia, Zichen~Cheng, and reviewers for suggestions on this work.
This work is supported by the Natural Science Foundation of China (NSFC No.62202105) and Shanghai Municipal Science and Technology Major Project (2021SHZDZX0103), General Program (No. 21ZR1403300), and Sailing Program (No. 21YF1402900).

\fi

\bibliographystyle{IEEEtran}

\ifx\hidemain\undefined
  \bibliography{assets/bibs/papers.bib,assets/bibs/labeling-systems.bib}
\fi

\ifx\hidemain\undefined
  \ifx\hidephoto\undefined
    \begin{IEEEbiography}[{\includegraphics[width=1in,height=1.25in,clip,keepaspectratio]{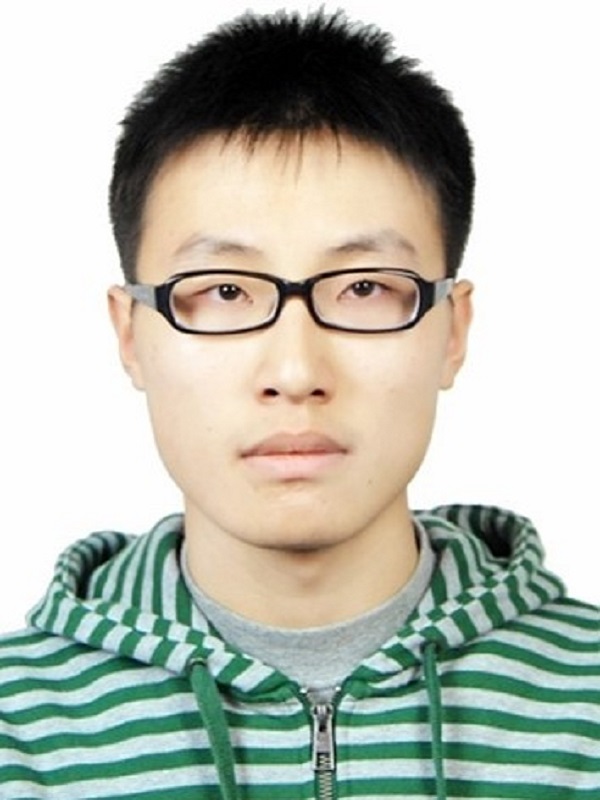}}]{Yu Zhang}
received a B.S. degree in Intelligence Science and Technology from Peking University in 2017.
Since then, he has been pursuing a Ph.D. at the Department of Computer Science, University of Oxford.
His research focuses on intelligent user interfaces and data visualization.
\end{IEEEbiography}

\begin{IEEEbiography}[{\includegraphics[width=1in,height=1.25in,clip,keepaspectratio]{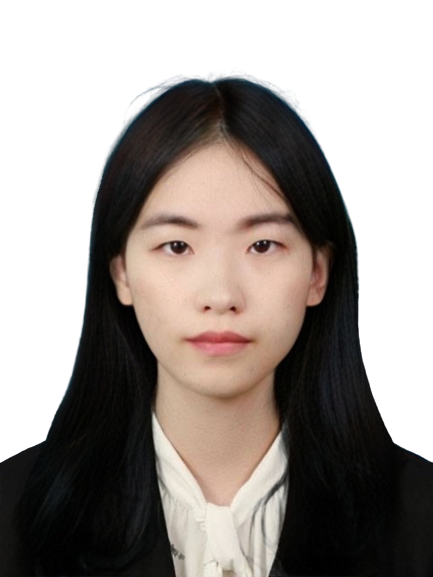}}]{Xinyi Zhao}
received a B.S. degree in Data Science and Big Data Technology from Fudan University.
She is currently pursuing a Master's degree in Data Science at Columbia University.
Her primary research interests lie in applying machine learning to enhance data visualization.
\end{IEEEbiography}

\begin{IEEEbiography}[{\includegraphics[width=1in,height=1.25in,clip,keepaspectratio]{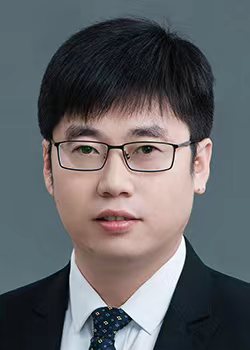}}]{Chongke Bi}
received the B.Sc. (Eng.) degree and the M.Sc. (Eng.) degree from Shandong University, China, in 2004 and 2007, respectively, and the Ph.D. (Sci.) degree from the University of Tokyo, Japan, in 2012.
After that, as a researcher at RIKEN, Japan, he focused on research in the field of visual analysis of big data on supercomputers from 2012 to 2016.
He is currently an associate Professor at Tianjin University.
His current research interests include visualization and high-performance computing.
\end{IEEEbiography}

\begin{IEEEbiography}[{\includegraphics[width=1in,height=1.25in,clip,keepaspectratio]{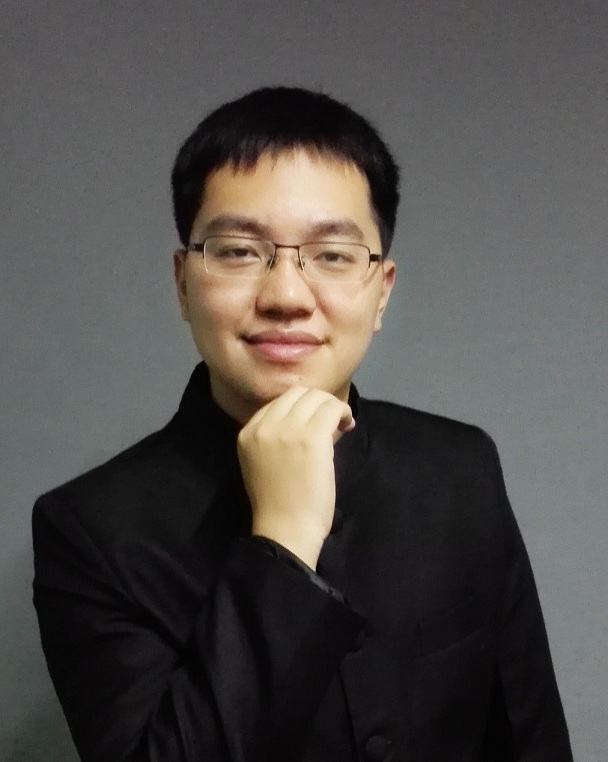}}]{Siming Chen}
is an Associate Professor at the School of Data Science, Fudan University.
Before this, he was a Research Scientist at Fraunhofer Institute IAIS in Germany.
He received his Ph.D. in computer science from Peking University.
His research interests are visualization and visual analytics, with an emphasis on human-AI collaboration, including LLM-driven visual analytics, social media, and autonomous driving visual analytics.
He has published 100 papers and over 40 in top conferences and journals, including IEEE VIS, IEEE TVCG, EuroVis, ACM CHI, UIST, CSCW, etc.
He served as multiple organizing chairs, committees, and reviewers.
He was awarded 10+ best paper/poster awards and honorable mention awards at multiple conferences.
For more information, please visit \url{http://simingchen.me}.
\end{IEEEbiography}

  \fi
\fi

\ifx\hideappendix\undefined
  \appendices
  \newpage
  \section{Proofs}
\label{sec:proofs}

This section provides the proofs of the mathematical derivations in the main text.

\subsection{Index of Difficulty of Curved Tunnel Passing}
\label{sec:proof-curved-tunnel-id}

\begin{mdframed}
    $ID$ is the index of difficulty of curved tunnel passing.

    \begin{itemize}[leftmargin=*]
        \item $ID = \int_0^{2r} \frac{dt}{w(t)}$
        \item $w(t) = W + 2r - 2\sqrt{r^2 - (r - t)^2}$
        \item $r \in \mathbb{R}^+$ and $W \in \mathbb{R}^+$
    \end{itemize}

    \noindent \textbf{One can show that} $\mathbf{ID = \frac{\frac{\pi}{2} + arcsin\frac{2r}{W + 2r}}{\sqrt{1 - \frac{4r^2}{(W + 2r)^2}}} - \frac{\pi}{2}}$
\end{mdframed}

\noindent $w(t)$ is symmetric with respect to $r$, i.e., $w(r - \delta) = w(r + \delta)$ for $\delta \in [0, r]$.

\noindent $\Rightarrow ID
    = 2\int_0^{r} \frac{dt}{w(t)}$

\noindent $\Rightarrow ID
    = 2\int_0^{r} \frac{dt}{W + 2r - 2\sqrt{r^2 - (r - t)^2}}$

\noindent Let $W_0 = W + 2r$.

\noindent $\Rightarrow ID
    = 2\int_0^{r} \frac{dt}{W_0 - 2\sqrt{r^2 - (r - t)^2}}$

\noindent Let $u = t - r$, then $du = dt$.

\noindent $\Rightarrow ID
    = 2\int_{-r}^{0} \frac{dt}{W_0 - 2\sqrt{r^2 - u^2}}$

\noindent $\Rightarrow ID
    = 2W_0 \int_{-r}^{0} \frac{du}{W_0^2 - 4r^2 + 4u^2} + 4 \int_{-r}^{0} \frac{\sqrt{r^2 - u^2}du}{W_0^2 - 4r^2 + 4u^2}$

\noindent Let
$A = \int_{-r}^{0} \frac{du}{W_0^2 - 4r^2 + 4u^2}$
and
$B = \int_{-r}^{0} \frac{\sqrt{r^2 - u^2}du}{W_0^2 - 4r^2 + 4u^2}$.

\noindent $\Rightarrow ID
    = 2W_0 A + 4 B$

\noindent In the following, we solve $A$ and $B$ respectively.

\noindent We first solve $A$:

\noindent Let $v = \frac{2u}{\sqrt{W_0^2 - 4r^2}}$, then $du = \frac{\sqrt{W_0^2 - 4r^2}}{2}dv$.

\noindent $\Rightarrow A
    = \int_{-\frac{2r}{\sqrt{W_0^2 - 4r^2}}}^{0} \frac{\frac{\sqrt{W_0^2 - 4r^2}}{2}dv}{W_0^2 - 4r^2 + (W_0^2 - 4r^2)v^2}$

\noindent $\Rightarrow A
    = \frac{1}{2\sqrt{W_0^2 - 4r^2}} \int_{-\frac{2r}{\sqrt{W_0^2 - 4r^2}}}^{0} \frac{dv}{1 + v^2}$

\noindent Note that $\frac{d(arctan(v))}{dv} = \frac{1}{1 + v^2}$.

\noindent $\Rightarrow A
    = \frac{1}{2\sqrt{W_0^2 - 4r^2}} (arctan0 - arctan \frac{-2r}{\sqrt{W_0^2 - 4r^2}})$

\noindent $\Rightarrow A
    = \frac{arcsin \frac{2r}{W_0}}{2\sqrt{W_0^2 - 4r^2}}$

\noindent Next, we solve $B$:

\noindent Let $u = rsin\alpha$, then $du = rcos\alpha d\alpha$.

\noindent $\Rightarrow B
    = \int_{-\frac{\pi}{2}}^{0} \frac{r^2cos^2\alpha d\alpha}{W_0^2 - 4r^2 + 4r^2sin^2\alpha}$

\noindent $\Rightarrow B
    = r^2 \int_{-\frac{\pi}{2}}^{0} \frac{d\alpha}{(W_0^2 - 4r^2)(1 + tan^2 \alpha) + 4r^2tan^2\alpha}$

\noindent Let $x = tan\alpha$, then $d\alpha = \frac{dx}{x^2+1}$.

\noindent $\Rightarrow B
    = r^2 \int_{-\infty}^{0} \frac{1}{W_0^2 - 4r^2 + W_0^2x^2} \times \frac{dx}{1 + x^2}$

\noindent Let $C = W_0^2 - 4r^2$ and $E = W_0^2$.

\noindent $\Rightarrow B
    = r^2 \int_{-\infty}^{0} \frac{1}{C + Ex^2} \times \frac{dx}{1 + x^2}$

\noindent Note that
$\frac{\frac{E}{E - C}}{C + Ex^2} - \frac{\frac{1}{E - C}}{1 + x^2} = \frac{1}{C + Ex^2} \times \frac{1}{1 + x^2}$.

\noindent $\Rightarrow B
    = r^2 \int_{-\infty}^{0} (\frac{\frac{E}{E - C}}{C + Ex^2} - \frac{\frac{1}{E - C}}{1 + x^2}) dx$

\noindent $\Rightarrow B
    = \frac{r^2}{E - C} \int_{-\infty}^{0} \frac{1}{\frac{C}{E} + x^2} dx - \frac{r^2}{E - C} \int_{-\infty}^{0} \frac{1}{1 + x^2} dx$

\noindent Let $F = \int_{-\infty}^{0} \frac{1}{\frac{C}{E} + x^2} dx$.

\noindent Let $y = \sqrt{\frac{E}{C}} x$, then $dx = \sqrt{\frac{C}{E}} dy$.

\noindent $\Rightarrow F
    = \int_{-\infty}^{0} \frac{1}{\frac{C}{E} + \frac{C}{E} y^2} \sqrt{\frac{C}{E}} dy$

\noindent $\Rightarrow F
    = \sqrt{\frac{E}{C}} \int_{-\infty}^{0} \frac{1}{1 + y^2} dy$

\noindent Plug in the value of $F$ to $B$.

\noindent $\Rightarrow B
    = \frac{r^2}{E - C} (\sqrt{\frac{E}{C}} - 1) \int_{-\infty}^{0} \frac{1}{1 + x^2} dx$

\noindent Note that $\frac{d(arctan(v))}{dv} = \frac{1}{1 + v^2}$.

\noindent $\Rightarrow B
    = \frac{r^2}{E - C} (\sqrt{\frac{E}{C}} - 1) (arctan0 - arctan(-\infty))$

\noindent $\Rightarrow B
    = \frac{r^2}{E - C} (\sqrt{\frac{E}{C}} - 1) \times \frac{\pi}{2}$

\noindent Plug in the value of $C$ and $E$ to $B$.

\noindent $\Rightarrow B
    = \frac{\pi}{8}(\sqrt{\frac{W_0^2}{W_0^2 - 4r^2}} - 1)$

\noindent Plug in the value of $A$ and $B$ to $ID$.

\noindent $\Rightarrow ID
    = 2W_0 A + 4B$

\noindent $\Rightarrow ID
    = \frac{W_0 arcsin \frac{2r}{W_0}}{\sqrt{W_0^2 - 4r^2}} + \frac{\pi}{2}(\sqrt{\frac{W_0^2}{W_0^2 - 4r^2}} - 1)$

\noindent $\Rightarrow ID
    = \frac{\frac{\pi}{2} + arcsin \frac{2r}{W_0}}{\sqrt{1 - \frac{4r^2}{W_0^2}}} - \frac{\pi}{2}$

\noindent Finally, plug in the value of $W_0$ to $ID$.

\noindent $\Rightarrow ID
    = \frac{\frac{\pi}{2} + arcsin \frac{2r}{W + 2r}}{\sqrt{1 - \frac{4r^2}{(W + 2r)^2}}} - \frac{\pi}{2}$
$\blacksquare$

\subsection{Upper and Lower Bounds of the Index of Difficulty of Dotted Fixed-width Tunnel Passing}
\label{sec:proof-dotted-tunnel-id-bounds}

\begin{mdframed}
    $ID$ is the index of difficulty of dotted fixed-width tunnel passing.

    \begin{itemize}[leftmargin=*]
        \item $ID$ is a function with regard to $k \in \mathbb{Z}^+$, $W \in \mathbb{R}^+$, $r \in \mathbb{R}^+$, $d \in \mathbb{R}_{\geq 0}$, and $m \in \mathbb{R}^+$.
        \item $ID = (k + 1) ID_1 + mk ID_2$
        \item $ID_1 = \frac{\frac{\pi}{2} + arcsin\frac{2r}{W + 2r}}{\sqrt{1 - \frac{4r^2}{(W + 2r)^2}}} - \frac{\pi}{2}$
        \item $ID_2 = log_2 (\frac{d}{W + 2r} + 1)$
    \end{itemize}

    \noindent \textbf{One can show that:}

    \begin{itemize}[leftmargin=*]
        \item \textbf{Monotonicity: $\mathbf{ID}$ is monotonically non-increasing with regard to $\mathbf{d}$ under the constraint that $\mathbf{D = 2r(k + 1) + dk}$ is a fixed value.}

        \item \textbf{Upper bound: $\mathbf{ID \leq \frac{D}{W}}$ and $\mathbf{\lim_{r \rightarrow 0^+} (ID \rvert_{d=0}) = \frac{D}{W}}$ under the constraint that $\mathbf{D = 2r(k + 1) + dk}$ is a fixed value.}

        \item \textbf{Lower bound: $\mathbf{ID \geq mk log_2(\frac{D - 2r(k + 1)}{k(W + 2r)} + 1)}$ and $\mathbf{\lim_{r \rightarrow 0^+} ID = mk log_2(\frac{D}{kW} + 1)}$.}
    \end{itemize}

\end{mdframed}

In the following, we prove the monotonicity, upper bound, and lower bound one by one.

\subsubsection{Monotonicity}

\textbf{Remark:}
Geometrically, it is intuitive that
under the constraint that $D = 2r(k + 1) + dk$ is a fixed value,
as $d$ increases, the points along the dotted tunnel are sparser,
and thus, $ID$ should not increase.
The following proves this intuition.

\noindent Plug in the constraint $k = \frac{D - 2r}{d + 2r}$ to $ID$.

\noindent $\Rightarrow ID = \frac{D + d}{d + 2r} (\frac{\frac{\pi}{2} + arcsin\frac{2r}{W + 2r}}{\sqrt{1 - \frac{4r^2}{(W + 2r)^2}}} - \frac{\pi}{2}) + m \frac{D - 2r}{d + 2r} log_2 (\frac{d}{W + 2r} + 1)$

\noindent Note that $ID_1 = \frac{\frac{\pi}{2} + arcsin\frac{2r}{W + 2r}}{\sqrt{1 - \frac{4r^2}{(W + 2r)^2}}} - \frac{\pi}{2}$ is irrelevant to $d$.

\noindent $\Rightarrow \frac{dID}{d(d)}
    = ID_1 \frac{d \frac{D + d}{d + 2r}}{d(d)} + m (\frac{d \frac{D - 2r}{d + 2r}}{d(d)} log_2 (\frac{d}{W + 2r} + 1) + \frac{D - 2r}{d + 2r} \times \frac{d log_2 (\frac{d}{W + 2r} + 1)}{d(d)})$

\noindent $\Rightarrow \frac{dID}{d(d)}
    = ID_1 \frac{2r - D}{(d + 2r)^2} + m (\frac{2r - D}{(d + 2r)^2} log_2 (\frac{d}{W + 2r} + 1) + \frac{D - 2r}{d + 2r} \times \frac{log_2 e}{W + 2r + d})$

\noindent Let
$A = \frac{2r - D}{(d + 2r)^2}$
and
$B = \frac{2r - D}{(d + 2r)^2} log_2 (\frac{d}{W + 2r} + 1) + \frac{D - 2r}{d + 2r} \times \frac{log_2 e}{W + 2r + d}$.

\noindent In the following, we show that $A \leq 0$ and $B \leq 0$.

\noindent Note that $D \in [4r + d, +\infty)$ where $r \in \mathbb{R}^+$ and $d \in \mathbb{R}_{\geq 0}$.

\noindent $\Rightarrow 2r - D < 0$.

\noindent $\Rightarrow A \leq 0$

\noindent Let $C = -log_2 (\frac{d}{W + 2r} + 1) + \frac{d + 2r}{d + W + 2r} log_2 e$.

\noindent To show $B \leq 0$ equals to showing $C \leq 0$.

\noindent Note that $\frac{dC}{d(d)} = log_2 e \frac{-2r - d}{(d + W + 2r)^2} \leq 0$.

\noindent $\Rightarrow$ $C$ is monotonically non-increasing with regard to $d$.

\noindent $\Rightarrow$ $C$ is maximized when $d$ is minimized, i.e., when $d = 0$, in which case $C = 0$.

\noindent $\Rightarrow$ $C \leq 0$

\noindent $\Rightarrow$ $B \leq 0$

\noindent $ID_1$ is the index of difficulty for curved tunnel passing.
It is easy to check that $ID_1 \geq 0$.

\noindent Note that $ID_1 \geq 0$, $A \leq 0$, $B \leq 0$, and $m \geq 0$.

\noindent $\Rightarrow \frac{dID}{d(d)} = ID_1 A + m B \leq 0$
$\blacksquare$

\subsubsection{Upper Bound}
\label{sec:upper-bound}

\textbf{Proof of the Limit:
    $\mathbf{\lim_{r \rightarrow 0^+} (ID \rvert_{d=0}) = \frac{D}{W}}$ under the constraint that $\mathbf{D = 2r(k + 1) + dk}$ is a fixed value.
}

\noindent Under the constraint that $D = 2r(k + 1) + dk$ is a fixed value, $ID = \frac{D + d}{d + 2r} (\frac{\frac{\pi}{2} + arcsin\frac{2r}{W + 2r}}{\sqrt{1 - \frac{4r^2}{(W + 2r)^2}}} - \frac{\pi}{2}) + m \frac{D - 2r}{d + 2r} log_2 (\frac{d}{W + 2r} + 1)$ as shown above.

\noindent $\Rightarrow ID \rvert_{d=0} = \frac{D}{2r} (\frac{\frac{\pi}{2} + arcsin\frac{2r}{W + 2r}}{\sqrt{1 - \frac{4r^2}{(W + 2r)^2}}} - \frac{\pi}{2})$

\noindent Let $\frac{2r}{W + 2r} = sin \alpha$ where $\alpha \in (0, \frac{\pi}{2})$.

\noindent $\Rightarrow ID \rvert_{d=0} = \frac{D}{W} (\frac{1}{sin \alpha} - 1) (\frac{\frac{\pi}{2} + \alpha}{cos \alpha} - \frac{\pi}{2})$

\noindent $\Rightarrow ID \rvert_{d=0} = \frac{D}{W} \frac{\frac{\frac{\pi}{2} + \alpha}{cos \alpha} - \frac{\pi}{2}}{\frac{sin \alpha}{1 - sin \alpha}}$

\noindent Note that when $r$ approaches $0^+$, $\alpha$ also approaches $0^+$.

\noindent $\Rightarrow \lim_{r \rightarrow 0^+}(ID \rvert_{d=0}) = \frac{D}{W} \lim_{\alpha \rightarrow 0^+} \frac{\frac{\frac{\pi}{2} + \alpha}{cos \alpha} - \frac{\pi}{2}}{\frac{sin \alpha}{1 - sin \alpha}}$

\noindent Note that when $\alpha$ approaches $0^+$, $\frac{\frac{\pi}{2} + \alpha}{cos \alpha} - \frac{\pi}{2}$ and $\frac{sin \alpha}{1 - sin \alpha}$ both approach $0^+$.
Apply L'Hôpital's rule.

\noindent $\Rightarrow \lim_{r \rightarrow 0^+}(ID \rvert_{d=0}) = \frac{D}{W} \lim_{\alpha \rightarrow 0^+} \frac{\frac{cos \alpha + (\frac{\pi}{2} + \alpha) sin \alpha}{cos^2 \alpha}}{\frac{cos \alpha (1 - sin \alpha ) + sin \alpha cos \alpha}{(1 - sin \alpha)^2}}$

\noindent $\Rightarrow \lim_{r \rightarrow 0^+}(ID \rvert_{d=0}) = \frac{D}{W}$
$\blacksquare$

\noindent \textbf{Proof of the Bound:
    $\mathbf{ID \leq \frac{D}{W}}$ under the constraint that $\mathbf{D = 2r(k + 1) + dk}$ is a fixed value.
}

\noindent As proved above, $ID$ is monotonically non-increasing with regard to $d$ under the constraint that $D = 2r(k + 1) + dk$ is a fixed value.
The valid range of $d$ is $[0, D - 4r]$.

\noindent $\Rightarrow ID \leq ID \rvert_{d=0}$

\noindent As shown above, $ID \rvert_{d=0} = \frac{D}{W} (\frac{1}{sin \alpha} - 1) (\frac{\frac{\pi}{2} + \alpha}{cos \alpha} - \frac{\pi}{2})$ where $\alpha \in (0, \frac{\pi}{2})$.

\noindent $\Rightarrow \frac{d (ID \rvert_{d=0})}{d \alpha} = \frac{D}{W} (\frac{-cos \alpha}{sin^2 \alpha} (\frac{\frac{\pi}{2} + \alpha}{cos \alpha} - \frac{\pi}{2}) + (\frac{1}{sin \alpha} - 1) (\frac{cos \alpha + (\frac{\pi}{2} + \alpha) sin \alpha}{cos^2 \alpha}))$

\noindent In the following, we show $\frac{d (ID \rvert_{d=0})}{d \alpha} \leq 0$ where $\alpha \in (0, \frac{\pi}{2})$.

\noindent Let $f(\alpha) = \frac{\pi}{2} cos^3 \alpha - (\frac{\pi}{2} + \alpha) cos^2 \alpha + sin \alpha cos \alpha - sin^2 \alpha cos \alpha + (\frac{\pi}{2} + \alpha) sin^2 \alpha - (\frac{\pi}{2} + \alpha) sin^3 \alpha$.

\noindent One can check that
if $f(\alpha) \leq 0$ when $\alpha \in [0, \frac{\pi}{2}]$,
then $\frac{d (ID \rvert_{d=0})}{d \alpha} \leq 0$ when $\alpha \in (0, \frac{\pi}{2})$.

\noindent Note that $\frac{d f(\alpha)}{d \alpha} = sin \alpha cos \alpha \times (-(\frac{3}{2} \pi + 2) cos \alpha - (\frac{3}{2} \pi + 3 \alpha) sin \alpha + (2 \pi + 4 \alpha))$.

\noindent Let $g(\alpha) = -(\frac{3}{2} \pi + 2) cos \alpha - (\frac{3}{2} \pi + 3 \alpha) sin \alpha + (2 \pi + 4 \alpha)$.

\noindent $\Rightarrow \frac{d g(\alpha)}{d \alpha} = (\frac{3}{2} \pi - 1) sin \alpha - (\frac{3}{2} \pi + 3 \alpha) cos \alpha + 4$

\noindent $\Rightarrow \frac{d^2 g(\alpha)}{d \alpha^2} = (\frac{3}{2} \pi - 4) cos \alpha + (\frac{3}{2} \pi + 3 \alpha) sin \alpha$

\noindent Note that $\frac{d^2 g(\alpha)}{d \alpha^2} > 0$ when $\alpha \in [0, \frac{\pi}{2}]$.

\noindent $\Rightarrow \frac{d g(\alpha)}{d \alpha}$ is monotonously increasing with regard to $\alpha$ when $\alpha \in [0, \frac{\pi}{2}]$.

\noindent Note that $\frac{d g(\alpha)}{d \alpha} \rvert_{\alpha = 0} = 4 - \frac{3}{2} \pi < 0$ and $\frac{d g(\alpha)}{d \alpha} \rvert_{\alpha = \frac{\pi}{2}} = 3 + \frac{3}{2} \pi > 0$.

\noindent $\Rightarrow$ There exist $\alpha_0 \in [0, \frac{\pi}{2}]$ such that
$\frac{d g(\alpha)}{d \alpha} \rvert_{\alpha = \alpha_0} = 0$, and
$\frac{d g(\alpha)}{d \alpha} < 0$ when $\alpha \in [0, \alpha_0)$, and
$\frac{d g(\alpha)}{d \alpha} > 0$ when $\alpha \in (\alpha_0, \frac{\pi}{2}]$.

\noindent $\Rightarrow$ There exist $\alpha_0 \in [0, \frac{\pi}{2}]$ such that
$g(\alpha)$ is monotonously non-increasing on $[0, \alpha_0]$, and
$g(\alpha)$ is monotonously non-decreasing on $[\alpha_0, \frac{\pi}{2}]$.

\noindent Note that $g(0) = \frac{\pi}{2} - 2 < 0$ and
$g(\frac{\pi}{2}) = \pi > 0$ and
$g(\alpha_0) \leq g(0) < 0$.

\noindent $\Rightarrow$ There exist $\alpha_1 \in [\alpha_0, \frac{\pi}{2}]$ such that
$g(\alpha_1) = 0$, and
$g(\alpha) < 0$ when $\alpha \in [0, \alpha_1)$, and
$g(\alpha) > 0$ when $\alpha \in (\alpha_1, \frac{\pi}{2}]$.

\noindent $\Rightarrow$ There exist $\alpha_1 \in [\alpha_0, \frac{\pi}{2}]$ such that
$\frac{d f(\alpha)}{d \alpha} \rvert_{\alpha = \alpha_1} = 0$, and
$\frac{d f(\alpha)}{d \alpha} < 0$ when $\alpha \in [0, \alpha_1)$, and
$\frac{d f(\alpha)}{d \alpha} > 0$ when $\alpha \in (\alpha_1, \frac{\pi}{2}]$.

\noindent $\Rightarrow$ There exist $\alpha_1 \in [\alpha_0, \frac{\pi}{2}]$ such that
$f(\alpha)$ is monotonously non-increasing on $[0, \alpha_1]$, and
$f(\alpha)$ is monotonously non-decreasing on $[\alpha_1, \frac{\pi}{2}]$.

\noindent Note that $f(0) = 0$ and $f(\frac{\pi}{2}) = 0$.

\noindent $\Rightarrow f(\alpha) \leq 0$ when $\alpha \in [0, \frac{\pi}{2}]$.

\noindent $\Rightarrow \frac{d (ID \rvert_{d=0})}{d \alpha} \leq 0$ when $\alpha \in (0, \frac{\pi}{2})$.

\noindent Note that $\alpha = arcsin \frac{2r}{W + 2r}$

\noindent $\Rightarrow \frac{d \alpha}{d r} = \frac{2W}{(W + 2r)^2\sqrt{1 - (\frac{2r}{W + 2r})^2}} \geq 0$

\noindent $\Rightarrow \frac{d (ID \rvert_{d=0})}{d r} = \frac{d (ID \rvert_{d=0})}{d \alpha} \times \frac{d \alpha}{d r} \leq 0$

\noindent $\Rightarrow ID \rvert_{d=0}$ is maximized when $r$ is minimized, i.e., when $r \rightarrow 0^+$.

\noindent As proved above, $\lim_{r \rightarrow 0^+}(ID \rvert_{d=0}) = \frac{D}{W}$.

\noindent $\Rightarrow ID \leq ID \rvert_{d=0} \leq \lim_{r \rightarrow 0^+}(ID \rvert_{d=0}) = \frac{D}{W}$
$\blacksquare$

\subsubsection{Lower Bound}
\label{sec:lower-bound}

\noindent \textbf{Proof of the Limit:
    $\mathbf{\lim_{r \rightarrow 0^+} ID = mk log_2(\frac{D}{kW} + 1)}$.
}

\noindent Recall $ID = (k + 1) (\frac{\frac{\pi}{2} + arcsin\frac{2r}{W + 2r}}{\sqrt{1 - \frac{4r^2}{(W + 2r)^2}}} - \frac{\pi}{2}) + m k log_2 (\frac{d}{W + 2r} + 1)$.

\noindent $\Rightarrow \lim_{r \rightarrow 0^+} ID = \lim_{r \rightarrow 0^+} (m k log_2 (\frac{d}{W} + 1))$

\noindent Note that $d = \frac{D - 2r(k + 1)}{k}$.
Plug in $d$ to $ID$.

\noindent $\Rightarrow \lim_{r \rightarrow 0^+} ID = \lim_{r \rightarrow 0^+} (m k log_2 (\frac{D - 2r(k + 1)}{kW} + 1))$

\noindent $\Rightarrow \lim_{r \rightarrow 0^+} ID = m k log_2 (\frac{D}{kW} + 1)$
$\blacksquare$

\noindent \textbf{Proof of the Bound:
    $\mathbf{ID \geq mk log_2(\frac{D - 2r(k + 1)}{k(W + 2r)} + 1)}$.
}

\noindent Recall $ID = (k + 1) (\frac{\frac{\pi}{2} + arcsin\frac{2r}{W + 2r}}{\sqrt{1 - \frac{4r^2}{(W + 2r)^2}}} - \frac{\pi}{2}) + m k log_2 (\frac{d}{W + 2r} + 1)$.

\noindent One can check that $\frac{\frac{\pi}{2} + arcsin\frac{2r}{W + 2r}}{\sqrt{1 - \frac{4r^2}{(W + 2r)^2}}} - \frac{\pi}{2} \geq 0$.

\noindent $\Rightarrow ID \geq m k log_2 (\frac{d}{W + 2r} + 1)$

\noindent Note that $D = 2r(k + 1) + dk$.

\noindent $\Rightarrow d = \frac{D - 2r(k + 1)}{k}$

\noindent Plug in $d$ to $m k log_2 (\frac{d}{W + 2r} + 1)$.

\noindent $\Rightarrow ID \geq m k log_2 (\frac{D - 2r(k + 1)}{k(W + 2r)} + 1)$
$\blacksquare$

  \section{Point Cloud Clustering}
\label{sec:point-cloud-clustering}

Semantic segmentation models classify points but do not differentiate semantic instances of the same category, e.g., points belonging to two vehicles are all categorized into the \category{vehicle} category.
By comparison, instance segmentation models distinguish individual instances of the same category, e.g., points belonging to two vehicles are differentiated as \category{$\mathtt{vehicle_1}$} and \category{$\mathtt{vehicle_2}$}.

As introduced in
\ifx\hidemain\undefined
    Sec.~\ref{sec:approach},
\else
    Sec. 4 in the main text,
\fi
our approach assumes the points in the point cloud are grouped into semantic instances.
When an instance segmentation model is applied to the point cloud data, the grouping is readily done.

When a semantic segmentation model is applied to the point cloud data, we adapt DBSCAN to obtain semantic instances.
Let $P_c$ denote the set of points categorized as $c$ by the semantic segmentation model.
For each label category $c$, we use DBSCAN to group points in $P_c$ into a set of clusters.
For example, for the \category{vehicle} category, the result is a collection of point sets \category{$\mathtt{vehicle_1}$}, \category{$\mathtt{vehicle_2}$}, ..., \category{$\mathtt{vehicle_n}$}.

The parameters of DBSCAN are determined as follows:

\begin{itemize}[leftmargin=*]
    \item \textbf{Neighborhood radius} $\mathbf{eps}$:
          The maximum distance for two points to be considered neighbors.
          We set $eps = 100 d_{op} \theta$ in our implementation.

          \begin{itemize}
              \item $d_{op}$ is the distance from the origin to a point $p$ in the 3D space.
              \item $\theta$ is the angular resolution of the LiDAR sensor.
          \end{itemize}

          Conventionally, DBSCAN's neighborhood radius is independent of the location of the clustered point.
          For the point clouds obtained with LiDAR sensors, the point density depends on the location, specifically, the distance from the origin.
          Thus, we let $eps$ depend on $d_{op}$.

    \item \textbf{Minimum neighborhood size} $\mathbf{minPts}$:
          The minimum number of points in the neighborhood for a point to be counted as a core point.
          In our implementation, we set $minPts = 10$.
\end{itemize}

\subsection{Considerations in Parameter Choices}

In the following, we introduce considerations in choosing the neighborhood radius $eps$.
We consider the case where the point cloud data is obtained with a LiDAR sensor.

\textbf{Setting neighborhood radius $\mathbf{eps \propto d_{op}}$:}
Assuming the LiDAR sensor scans evenly in all directions, the density of scanned points is proportional to the distance from the origin (i.e., the place where the LiDAR sensor is placed).
Thus, we set $eps$ proportional to $d_{op}$.

\textbf{Setting neighborhood radius $\mathbf{eps \propto \bm{\theta}}$:}
The density of scanned points is proportional to $\theta$.
Thus, we set $eps$ to be proportional to $\theta$.

\textbf{Estimating angular resolution $\bm{\theta}$:}
For some point cloud datasets, the angular resolution $\theta$ is readily available.
When the angular resolution is unknown, we estimate it from the point cloud data.
For each point $p$, we find its nearest neighbor $q$.
Let $d_{pq}$ denote the distance between $p$ and $q$.
Let $\theta_p$ denote the angular resolution at point $p$.
We estimate $\theta_p$ as the angle formed by the three points, $p$, origin, and $q$.
$\theta_p$ is typically small, and thus $\theta_p \approx \sin(\theta_p)$.
Therefore, $\theta_p \approx \frac{d_{pq}}{d_{op}}$.
We estimate $\theta$ as the median of $\theta_p$ over all points $p$, i.e., $ \theta = \underset{p}{median}(\theta_p)$.

\fi

\end{document}